\documentclass{article}

\usepackage{amsfonts,amsmath}
\usepackage{fullpage}
\usepackage{url} 
\usepackage{newfloat}
\usepackage{multirow}
\usepackage{framed}
\DeclareFloatingEnvironment[
    fileext=los,
    listname=List of Goals,
    name=Optimization Goal,
    placement=tbhp,
]{goal}
\usepackage{breqn}

\usepackage{mathrsfs}
\usepackage{color}
\usepackage{algorithm}
\usepackage{algpseudocode}

\newcommand{\fullversion}[1]{#1}
\newcommand{\confversion}[1]{}
\newcommand{\cvspace}[1]{\confversion{\vspace{#1}}}

\usepackage{subcaption}
\newtheorem{theorem}{Theorem}

\def \QED {\hfill{$\Box$}}

\newenvironment{proofof}[1]{\noindent {\em Proof of #1.  }}{\QED}

\newenvironment{remindertheorem}[1]{\medskip \noindent {\bf Reminder of Theorem #1.  }\em}{}

\usepackage{graphicx}
\begin{document}

\newcommand{\pwdspace}{\mathscr{P}}
\newcommand{\padv}{P_{adv}}
\newcommand{\padvB}{P_{adv,B}}
\newcommand{\pdet}{P_{det}}
\newcommand{\pdetB}{P_{det,B}}
\newcommand{\pwdratio}{\frac{B}{k|\pwdspace|}}
\newcommand{\pwdpartratio}[1]{\frac{B_{#1}}{|\pwdspace|}}
\newcommand{\prob}[1]{\tilde p_{#1}}
\newcommand{\hprob}[1]{\Pr[E_{#1}]}
\newcommand{\pred}[1]{P_{#1}}
\newcommand{\upred}[1]{P_{u_{#1}}}
\newcommand{\invpwdratio}{\frac{k|\pwdspace|}{B}}
\newcommand{\predevent}[1]{\{\text{select } P_{#1}\}}
\newcommand{\stopcond}{\mathcal{S}}
\newcommand{\outputspace}{\mathcal{O}}
%Let's leave the \selpred macro without a space. We can insert a space manually,
% For example, $\selpred$ will leave a space or \selpred~will leave a space. 
\newcommand{\selpred}{\textbf{SelPreds}}
\newcommand{\clientcash}{Client-CASH}
\newcommand{\reproduce}{\textbf{Reproduce }}
\newcommand{\outputsize}{|\outputspace|}
\newcommand{\coeffone}{\frac{\ell_1}{\ell_1-1}}
\newcommand{\coefftwo}{\frac{\ell_1\ell_2}{(\ell_1-1)(\ell_2-1)}}
\newcommand{\advstrat}[1]{\mathcal{A}_{#1}}
\newcommand{\hash}{\mathbf{H}}
\newcommand{\alphak}{\frac{C_{srv}}{k\cdot C_H}}

\newcommand{\padvformula}{
\pwdratio \max\limits_{\mathbf{b} \in F} \left\{
\hprob{1}b_1 + \sum\limits_{i=2}^n \hprob{i}b_i \prod\limits_{j=1}^{i-1} \left( \frac{\ell_j}{\ell_j-1} \right)
\right\}
}

\newcommand{\padvfunction}{
\hprob{1}b_1 + \sum\limits_{i=2}^n \hprob{i}b_i \prod\limits_{j=1}^{i-1} \left( \frac{\ell_j}{\ell_j-1} \right)
}

%cinnamon :D
\definecolor{cinnamon}{rgb}{0.82, 0.41, 0.12}

%editing comments
\newcommand{\jnote}[1]{{\color{red}{\bf [Jeremiah: #1]}}}
\newcommand{\anote}[1]{{\color{cinnamon}{\bf [Ani: #1]}}}

\newcommand{\checkpoint}{
{\color{blue}{\bf [Checkpoint: Have revised paper until this point. ]}}
}

\newcommand{\progress}{
{\color{blue}{\bf [PROGRESS: Currently working on this section. ]}}
}

\newcommand{\cut}[1]{}

%\mainmatter

\title{\clientcash: Protecting Master Passwords against Offline Attacks}
%\titlerunning{\clientcash}

%\numberofauthors{2}
\author{Jeremiah Blocki \\ Microsoft Research \\ jblocki@microsoft.com
\and
Anirudh Sridhar \\
Carnegie Mellon University \\
asridha1@andrew.cmu.edu
} 
\maketitle

% !TEX root = ASIA_CCS_PWD_SYSTEM_2016.tex

\begin{abstract} 
Offline attacks on passwords are increasingly commonplace and dangerous. An offline adversary is limited only by the amount of computational resources he or she is willing to invest to crack a user's password. The danger is compounded by the existence of authentication servers who fail to adopt proper password storage practices like key-stretching. Password managers can help mitigate these risks by adopting  key stretching procedures like hash iteration or memory hard functions to derive site specific passwords from the user's master password on the client-side. While key stretching can reduce the offline adversary's success rate, these procedures also increase computational costs for a legitimate user. Motivated by the observation that most of the password guesses of the offline adversary will be incorrect, we propose a client side cost asymmetric secure hashing scheme (\clientcash). \clientcash~randomizes the runtime of client-side key stretching procedure in a way that the expected computational cost of our key derivation function is greater when run with an incorrect master password. We make several contributions. First, we show how to introduce randomness into a client-side key stretching algorithms through the use of  halting predicates which are selected randomly at the time of account creation. Second, we formalize the problem of finding the optimal running time distribution subject to certain cost constraints for the client and certain security constrains on the halting predicates. Finally, we demonstrate that \clientcash~can reduce the adversary's success rate by up to $21\%$. These results demonstrate the promise of the \clientcash~mechanism. %We speculate that with further analysis, we can obtain even higher reductions.
\end{abstract}
\section{Introduction}
%\vspace{-0.30cm}
\label{sec:Introduction}
% !TEX root = ASIA_CCS_PWD_SYSTEM_2016.tex
Passwords are often the primary entryway to access a user's confidential information on a website, and are thus a focus of attention for attackers. Offline attacks against passwords are particularly powerful. An offline adversary has access to the cryptographic hash of a user's password and can check a vast number of password possibilities without interacting with the authentication server. This adversary is only restricted by the computational resources that he is willing to invest into breaching each account. Offline attacks are increasingly commonplace due to recent high-profile security breaches at organizations like LinkedIN, Sony, eBay, and Ashley Madison\footnote{For example, see \url{ http://www.privacyrights.org/data-breach/} (Retrieved 9/1/2015).}. 

Several factors contribute to the danger of offline attacks. First, users struggle to remember high entropy passwords for multiple accounts. Second, many organizations had failed to implement proper password storage techniques at the time they were breached. Finally, password cracking resources (e.g., hardware and password dictionaries) continue to improve allowing an adversary to mount cheaper and more effective attacks. 

Client-side password management tools (e.g., PwdHash~\cite{ross2005stronger}) allow the user to generate multiple passwords from one master password and apply secure procedures like key stretching or salting on the client-side. While password managers offer many benefits they also introduce a single point of failure (i.e., the master password) through which the adversary could attack to gain access to all of a user's accounts. A user's master password is not necessarily immune to offline attacks just because a password manager is used. The recent breach of LastPass\footnote{For example, see \url{https://blog.lastpass.com/2015/06/lastpass-security-notice.html/} (Retrieved 9/1/2015).} highlights this dangerous possibility.  Thus, key stretching procedures (e.g., hash iteration, memory hard functions) are recommended to mitigate security risks in the event of an offline attack. However, key-stretching increases costs for the honest party as well as the adversary. It is thus desirable to make authentication costs asymmetric so that (on average) a user authenticating with a correct password will incur lower costs than an offline adversary attempting to check an incorrect password guess. Previous work by Manber~\cite{manber1996simple} and by Blocki and Datta~\cite{blockiCASH2015} achieves this goal of cost-asymmetric key-stretching, but these solutions are only appropriate for server-side key stretching (see discussion in Section \ref{sec:Related Work}). 

\cut{\anote{Can cut down this section. }Several factors contribute to the danger of offline attacks. First, users often struggle to remember high entropy passwords for multiple different accounts. As a result many users tend to either select passwords with relatively low entropy and/or reuse the same password for multiple accounts. Second, many organizations had failed to implement proper password storage practices like key-stretching and salting at the time their servers were breached. Finally, password cracking resources (e.g., GPUs and password dictionaries) continue to improve. The result is that password cracking is often relatively cheap and that the attacker is often rewarded with access to multiple user accounts. Users can combat such an attack by using client-side password management tools (e.g., PwdHash~\cite{ross2005stronger}) which allow the user to generate multiple passwords from one master password. These tools allow the user to focus on memorizing one higher entropy password and they can ensure that proper techniques like key-stretching and salting are applied on the client-side before the derived password is ever sent to an authentication server.} 

\cut{However, such password management tools ~\cite{ross2005stronger} also introduce a single point of failure (the master password) which the adversary can attack to gain access to all of a user's accounts. The recent breach of LastPass\footnote{For example, wee \url{https://blog.lastpass.com/2015/06/lastpass-security-notice.html/} (Retrieved 9/1/2015).} highlights this dangerous possibility - a user's master password is not necessarily immune to offline attacks. Key stretching procedures such as iterated hashing are recommended to mitigate risks from an offline attack. However, key-stretching symmetrically increases costs for the honest party as well as the adversary. It is thus desirable to make authentication costs asymmetric so that the cost of user verification is less than the adversary's costs when attempting to breach the account with an incorrect password.}

\cut{While password management tools like PwdHash~\cite{ross2005stronger} offer many benefits to the users they also introduce a single point of failure (the master password) through which the adversary could attack to gain access to all of a user's accounts. The recent breach of LastPass\footnote{For example, see \url{https://blog.lastpass.com/2015/06/lastpass-security-notice.html/} (Retrieved 9/1/2015).} highlights this dangerous possibility. A user's master password is not necessarily immune to offline attacks just because a password manager is used. Thus, key stretching procedures such as using an iterated hash are recommended to mitigate risks in the event of an offline attack. However, key-stretching increases costs for the honest party as well as the adversary. It is thus desirable to make authentication costs asymmetric so that a user authenticating a correct password will have lower costs than an offline adversary attempting to breach the account with an incorrect password. }

\cvspace{-0.18cm}
\paragraph{Contributions} In this work we present \clientcash~a client-side key-stretching algorithm which achieves the goal of asymmetric costs, and we demonstrate that this system can protect user passwords from offline attacks. The core idea behind \clientcash~is to randomize the runtime of the client-side key stretching so that on average, the cost of verifying a correct password is smaller than the cost of rejecting an incorrect password. This is achieved by the use of halting predicates $P:\{0,1\}^*\rightarrow\{0,1\}$, which tells us when to stop hashing the password (e.g., stop after $t$ rounds of hash iteration if $P(\hash^t(pwd)) = 1$). These halting predicates are chosen randomly at the time of account creation and are stored by the client (e.g., on a local computer or on the cloud). 

Because these halting predicates are stored by the client we must take care to ensure that the predicates themselves do not leak too much information about the user's password. The key challenge is to select the predicates in a way that satisfies two seemingly conflicting requirements: 1) the halting predicates should induce a cost-asymmetry in the validation of correct/incorrect passwords, and 2) the adversary should not be able learn much about any user's master password even if he observes these halting predicates. We borrow ideas from differential privacy~\cite{dwork2006calibrating,mcsherry2007mechanism} to satisfy both of these conflicting requirements.  

We formalize the properties of \clientcash~and give an example of a system that satisfies these properties based on the Exponential Mechanism~\cite{mcsherry2007mechanism}, a powerful tool in differentially private analysis. We additionally formalize the problem of minimizing the percentage of passwords cracked by the adversary as an optimization over the parameters of the system and show that it can be solved efficiently as a linear program. In this work we analyze the security of \clientcash~systems that use up to two iterations of an underlying hash function $\mathbf{H}_k$ for password key stretching (the two-round case) or up to three iterations of $\mathbf{H}_k$ for key stretching (the three-round case). Here, the   parameter $k$ specifies the cost of the underlying hash function (e.g., BCRYPT~\cite{bcrypt} uses hash iteration to control costs and SCRYPT~\cite{percival2012scrypt} uses a memory hard function to control costs). If we let $\hprob{i,pwd}$ denote the probability that \clientcash~terminates after the $i$'th round given the correct password $pwd$ as input then the expected cost to verify a correct password guess is $k\sum_i i\cdot \hprob{i,pwd}$. 

We compare \clientcash~to deterministic key-stretching techniques $\mathbf{H}_{k'}$ with equivalent costs $k' = k\sum_i i\cdot \hprob{i,pwd}$  to \clientcash~by looking at the probability that an offline adversary with a finite computational budget $B$ could crack the user's password --- we use $\padvB$ (resp. $\pdetB$) to denote the adversary's success rate against \clientcash~(resp. deterministic key-stretching).  For the two-round case, we obtain a reduction $\pdetB-\padvB$ in passwords breached of up to $12\%$, and up to $21\%$ in the three round case. Although we only show results for 2 and 3 round systems, we formulate the essential groundwork for analysis of systems with more rounds. From the significant decreases in passwords breached from two round to three round systems, we might expect that introducing more rounds will decrease the number of passwords breached in an offline attack even further. 

In our analysis we make no assumptions on the password storage practices of the authentication server. If the authentication server adopts techniques like key stretching and salting then this will only make the adversary's task harder. However, we do assume that we face an optimal offline adversary with a finite computational budget $B$, and that users choose passwords uniformly at random. Because the later assumption does not hold for general users we only recommend our solution to users who choose passwords (nearly) uniformly at random. However, we argue that this assumption is plausible for many of the security conscious users that would opt to use a client-side key stretching algorithm (finding \clientcash~distributions that are optimal for protecting non-uniform passwords is an important direction for future work). 
\cvspace{-0.18cm}
\paragraph{Overview of \clientcash} When Alice creates an account on server $W$ using the master password $pwd_A$ our client-side application first selects a sequence of halting predicates $o_W = (\upred{1}$,$\ldots$, $\upred{n-1})$ using a randomized function $\selpred{(pwd_A)}$. Afterwards our application sends the message  $(Alice$, $H)$  to the server where $H=\hash_k^{\stopcond_W(pwd_A) }(pwd_A)$. Here, $\hash_k$ denotes a collision resistant hash function which costs $k$ work units to compute one time, $\hash_k^i$ denotes the hash function iterated $i$ times  and $\stopcond_W(pwd_A)$ denotes the stopping time for password $pwd_A$ given by the halting predicates $o_W$ --- that is the smallest number $i \geq 1$ for which $\upred{i}\big(\hash_k^{i}(pwd_A)\big)=1$. During authentication, Alice recomputes the derived password $H=\hash_k^{\stopcond_W(pwd_A)} (pwd_A)$, sends $(Alice,H)$ to the server and gains access to her account. An incorrect password guess $p_g \neq p_A$ would be rejected with high probability since $\hash_k$ is collision resistant.

%\vspace{-0.40cm}
\section{Related Work}
%\vspace{-0.40cm}
\label{sec:Related Work}
% !TEX root = ASIA_CCS_PWD_SYSTEM_2016.tex

\noindent{\bf Halting Puzzles.} At a high level our use of halting predicates is similar to Boyen's~\cite{boyen2007halting} halting puzzles. In Boyen's solution the chosen halting predicate will never return `halt' unless we run the key derivation function with the correct password so the key derivation algorithm never halts (or only halts after the maximum possible number of rounds). The key difference between our work and the work of Boyen~\cite{boyen2007halting} is that we carefully bound the amount of information that the chosen halting predicate(s) can leak about the user's password. Thus, unlike \cite{boyen2007halting}, we can ensure that an adversary who only breaches the client will not be able to execute an offline attack against the user's password.

\noindent {\bf Cost-Asymmetric Server-Side Key Stretching. }Manber~\cite{manber1996simple} proposed the use of hidden salt values (e.g., `pepper') to make it more expensive to reject incorrect passwords. Blocki and Datta~\cite{blockiCASH2015}  refined this idea using game theoretic tools. While our work closely follows the work of Blocki and Datta~\cite{blockiCASH2015}, we stress that neither work ~\cite{blockiCASH2015,manber1996simple} addresses the issue of client side key-stretching. In both of these schemes the authentication server selects a secret salt value $t \in \{1,\ldots,m\}$ (e.g., ``pepper") and stores the cryptographic hash $\hash(pwd,t)$ --- the value $t$ is not stored on the authentication server. An adversary would need to compute the hash function $m$ times in total to reject an incorrect password, while the authentication server will need to compute it {\em at most} $(m+1)/2$ times on average to verify a correct password guess because it can halt immediately after it finds the correct value of $t$. This approach is not suitable for client-side key-stretching because we would produce $m$ different derived keys, but the client program would not know which one is correct ---  neither of these value $t$ or $\hash(pwd,t)$ should be stored on the client. Since we are performing key-stretching on the client-side we need to ensure that the final derived password that is sent to the authentication server is consistent among different authentication sessions.

\noindent {\bf Password Management Software.} Password managers like PwdHash~\cite{ross2005stronger} allow the user to generate multiple passwords from one master password. PwdHash uses a public key-derivation function to (re)generate each of the user's passwords from a single master password. Since the key-derivation function is public an adversary who breaks into any of the third party authentication servers could still execute an offline attack against the user's master password. By contrast, password managers like KeePass~\cite{reichl2013keepass} store an encrypted password vault on the client and are not necessarily vulnerable in the previous scenario because the adversary would not have the password vault. The vault, which contains all of the user's passwords, is encrypted with the user's master password. However, any adversary who breaks into the client and steals a copy of this vault could execute an offline attack against the user's master password. Commercial applications like LastPass rely on a trusted server to derive passwords from the user's master password. Unlike PwdHash and KeePass these commercial applications are typically not open source so it is not always possible for independent researchers to verify their security properties. In theory these password managers could be designed so that an adversary would need to break into multiple servers (e.g., at LinkedIn and LastPass) before he can mount an offline attack on the user's master password. However, the recent breach at LastPass\footnote{See \url{https://blog.lastpass.com/2015/06/lastpass-security-notice.html/} (Retrieved 9/1/2015).} demonstrates that we cannot rule out this dangerous possibility. Similarly, Client-CASH is designed so that the adversary would need to breach both the client computer and a third party authentication server to mount an offline attack. 

\noindent {\bf Deterministic Key Stretching Techniques.} Advances in computing hardware (e.g., GPUs~\cite{kim2011gpu}, ASiCs~\cite{durmuthpassword}) make offline attacks increasingly dangerous. An offline adversary can often try millions of password guesses per second. Morris and Thompson~\cite{morris1979password} proposed the idea of key-stretching to make the hash function more expensive to evaluate so that an offline attack is more expensive for the adversary. Other defenses (e.g., distributing the storage/computation of cryptographic hash values so that an adversary who only breaches one server does not learn anything about the user's password \cite{brainard2003new} and ~\cite{camenisch2012practical}) require multiple dedicated authentication servers. Finding good key-stretching techniques is an active area of research\footnote{For example, the Password Hashing Competition (\url{https://password-hashing.net/index.html}) was developed to encourage the development of alternative password hashing schemes (e.g., \cite{Argon2,forler2013catena})}. Hash iteration (e.g., PBKDF2~\cite{kaliski2000pkcs}, BCRYPT~\cite{bcrypt}) alone is often viewed as an insufficient key-stretching technique because an offline adversary can often significantly reduce costs by building customized hardware to evaluate the iterated hash function. While computational speeds may vary greatly between different devices, memory latency speeds are relatively consistent~\cite{DGN03}. Thus, modern password hash functions like   SCRYPT~\cite{percival2012scrypt} or Argon2~\cite{Argon2} typically use memory hard functions~\cite{DGN03} for key-stretching purposes. Our work is largely orthogonal to these lines of research. In particular, we stress that \clientcash is compatible with both forms of key-stretching (hash iteration and memory hard functions).

\noindent{\bf Password Alternatives.} Although researchers have been working on alternatives to text-passwords (e.g., graphical passwords\cite{Jermyn:1999:DAG:1251421.1251422,brostoff2000passfaces,biddle2012graphical} or biometrics~\cite{bonneau2012quest}) text-passwords are likely to remain entrenched as the dominant form of authentication for many years~\cite{bonneau2012quest}. While we focus on text passwords in this paper we stress that the applications of \clientcash~are not necessarily limited to text passwords. Client-side key stretching is a valuable primitive that could be used to protect any lower entropy secret whether that secret is a text password, a graphical password or a biometric signal.

\cut{
\noindent {\bf Multiple Servers.} If we have multiple authentication servers we it is possible to distribute the storage of our cryptographic hash value across these servers so that an adversary who breaches one authentication server learns nothing about the user's password\endnote{However, an adversary who breaches all of the authentication servers will still be able to mount an offline attack.}. This approach was explored by Brainard et al.~\cite{brainard2003new} and more recently by Camenisch et al.~\cite{camenisch2012practical}. Juels and Rivest~\cite{rivestHoneywords} proposed to use the second server in a different way. They proposed to have the primary authentication server store the cryptographic hashes of honeywords (fake passwords) along with the cryptographic hashes of the user's real password, while an auxiliary server keeps track of which passwords are real or fake. This approach allows us to detect the offline attacker attacks when he starts to use the cracked passwords because the adversary will not always be able to tell which passwords are fake unless he also breaches the auxiliary server. The downside is that these defenses both require a second authentication server. 

\noindent {\bf AI Defenses against Offline Attacks.} Another line of research has sought to use hard artificial intelligence problems to protect passwords against offline attacks~\cite{canetti2006mitigating,poshExperiment,blockiGOTCHA}. The basic idea is to generate a hard artificial intelligence puzzle from the user's password and require the user to solve the puzzle to authenticate. The benefit of this extra work is that the offline adversary will need to solve a different hard AI puzzle to check each different password guess. \cut{The solution of Canetti et al. required filling a storage device on the authentication server with unsolved CAPTCHA puzzles. The downside to this approach is that an adversary who could afford to pay humans to solve these CAPTCHAs would be able to completely circumvent the scheme~\cite{motoyama2010re}. Blocki et al.~\cite{blockiGOTCHA} proposed a similar scheme called GOTCHA based on randomly generated inkblot images . Their scheme

They proposed filling a storage device with unsolved CAPTCHA~\cite{captcha} puzzles. To verify each password guess during an offline attack the adversary would need to solve a random CAPTCHA from the storage device. One downside to this approach is that the legitimate user must solve a CAPTCHA puzzle every time he wishes to authenticate. Another downside is that an adversary could completely circumvent the AI challenge by paying humans so solve every CAPTCHA puzzle on the storage device~\cite{motoyama2010re}. While this attack would be expensive, it could be worthwhile if it allowed the adversary to crack millions of passwords. Blocki et al.~\cite{blockiGOTCHA} proposed a similar scheme called GOTCHA, which used randomly generated inkblot images as part of the AI challenge. While their scheme does not rely on pregenerated challenges, the usability costs of this scheme appear to prohibitive for most applications.}

\noindent {\bf Measuring Password Strength.} Guessing-entropy~\cite{shannon1959mathematical,massey1994guessing}, $\sum_{i=1}^n i\times p_i$, measures the average number of guesses needed to crack a single password. Guessing-entropy and Shannon-entropy are known to be poor metrics for measuring password strength\endnote{Guessing-entropy could be high even if half of our users choose the same password ($p_1 = 0.5$) as long as the other half of our users choose a password uniformly at random from $\PasswordSpace$ $\left(p_2 = \ldots = p_n = \frac{2}{n-1}\right)$.}. While minimum entropy, $H_{\infty}=-\log p_1$, is a pessimistic security measurement it is a useful tool for worst-case analysis and it has been used to estimate  the fraction of accounts that could be cracked in an online attack~\cite{blockiPasswordComposition}. 

 Boztas~\cite{boztas1999entropies} proposed a metric called $\beta$-guesswork, which measured the success rate for an adversary with $\beta$ guesses per account $\sum_{i=1}^{\beta} p_i$. We use a similar formula for computing the budget-$B$ adversary success rate against our CASH mechanism --- the key difference is that the adversary must guess the random value $t_u$ as well as the user's password $pwd_u$. Pliam's proposed a similar metric called $\alpha$-guesswork~\cite{pliam2000incomparability}. This metric measure how many guesses the adversary would need to achieve success rate $\alpha$. 

Bonneau~\cite{bonneau2012science} created a system for sampling from the Yahoo! password distribution. This system allows one to gain empirical data about the frequency distribution of passwords without revealing the passwords themselves (e.g., for each user $u$ the system collected the value $\mathbf{H}\left(pwd_u,s \right)$ using a long salt value that was discarded after the experiment). Bonneau~\cite{bonneau2012science} collected and analyzed approximately $70$ million Yahoo passwords using this framework. We note that a similar sampling framework could be adopted to help compute the optimal CASH distribution. Our algorithms for finding the optimal CASH distribution only required the frequency distribution over passwords.

\noindent {\bf Encouraging Users to Memorize Stronger Passwords.} A separate line of research has focused on helping users memorize stronger passwords using various mnemonic techniques and/or rehearsal techniques (e.g.,~\cite{blockiNaturallyRehearsingPasswords,BS14,usabilitystudy:xkcd,Hertzum:2006:minimalFeedbackHintsforPasswords}). Password managers seek to minimize user burden by using a single password to generate multiple passwords~\cite{ross2005stronger}. These password managers often use client-side key stretching to produce each password. While CASH is a useful tool for server-side key stretching, our current version of CASH is not appropriate for client-side key stretching because the procedure is not deterministic. An interested direction for future research is determining whether similar ideas could be applied to obtain cost optimal client-side key stretching algorithms.
}

%\vspace{-0.4cm}
\section{Description of the Mechanism}
%\vspace{-0.4cm}
\label{sec:Description of the Mechanism}
% !TEX root = ASIA_CCS_PWD_SYSTEM_2016.tex
In this section we introduce the \clientcash~ mechanism and describe the account creation and authentication protocols. In our presentation we will use $\hash$ to denote a cryptographic hash function (e.g., SHA256 or Blake2b) and we will use $C_H$ to denote the cost of computing $\hash$ one time. We will also use $\hash_k$ to denote a deterministic hash function that is $k$ times more expensive to compute than $\hash$ (i.e., $C_{H_k} = k\cdot C_H$). This might be achieved by hash iteration~\cite{bcrypt,kaliski2000pkcs} or by the adoption of memory hard functions~\cite{percival2012scrypt}. We also use $\pwdspace$ to denote the set of passwords a user can pick. 
%We also informally describe properties of the system parameters which we use as the basis to design the specifics of the mechanism in later sections. 

\noindent{\bf Account Creation.} When a user $u$ creates an account $a$ with a master password $pwd_u \in \pwdspace$ \clientcash~will execute the following steps: First, the client will run a randomized algorithm $\selpred(pwd_u)$ to obtain a sequence of $n-1$ halting predicates $o_u  = \big(\upred{1},\ldots,\upred{n-1}\big)$. Here, a halting predicate $\upred{}:\{0,1\}^*\rightarrow\{0,1\}$ is simply a function that will tell us when to halt the key-derivation process. Second, the client will then store the tuple $\left(a,u,s_u,o_u\right)$, where $s_u \gets \mathbf{Unif}\left(\{0,1\}^L\right)$ is a random $L$-bit salt value. The client will then run the algorithm $\reproduce$ (described below) to derive the password for account $a$.  

We intentionally omit the workings of $\selpred$  and treat it as a black box for now. However, we stress that outcome $o_u$ selected by the randomized algorithm $\selpred(pwd_u)$ may depend both on the master password $pwd_u$ and on a security parameter $\epsilon$ which bounds the amount of information the outcome $o_u$ might leak about $pwd_u$.  In later sections we will show how to construct a randomized algorithm $\selpred$ which minimizes the adversary's success rate $\padvB$ subject to certain security and cost constraints.  The account creation protocol is described formally in Algorithm \ref{alg:CreateAccount}.

\begin{algorithm} 
\centering
\begin{algorithmic}
\State {\bf Input:} account name $a$, username $u$, password $pwd_u$; random bit strings $r_1, r_2$
\State {\bf System Parameters: }  rounds $n$, $\outputspace$, iterations $k$, $\epsilon$,  $L$ 
%\State $params \gets \left( n,\outputspace,\epsilon\right)$
\State $o_u \gets \selpred(pwd_u; r_1)$
\State $key \gets (u,a)$; $s_u \gets \mathbf{Unif}\left(\{0,1\}^L;r_2\right)$; $value \gets (o_u,s_u)$
\State $\mathbf{StoreOnClient}(key,value)$
\State $H \gets \reproduce(u,pwd_u,s_u,n)$ 
\State {\bf SendToServer}$(u,H)$
\end{algorithmic}
 \caption{Create Account (Client Side) }
\label{alg:CreateAccount}
\end{algorithm}

\noindent {\bf Authentication.} When the user $u$ attempts to access the account $a$ with the password guess $pwd_g \in \pwdspace$, the client first locates the record $\left(a,u,s_u,o_u\right)$ on the client. Then we execute the algorithm \reproduce function to derive the password $\hash_k^{\stopcond(pwd_g,o_u)}(pwd_g,s_u)$ for account $a$.

\begin{algorithm}[tbh]
\centering
\begin{algorithmic}
\State {\bf Input:}  account name $a$, username $u$, password $pwd_g$
\State $key \gets (u,a)$;  $value \gets \textbf{FindClientRecord}(key)$
\State {\bf If $value = \emptyset$ then return} ``Account does not exist."
\State $(o_u,s_u, n,\outputspace,\epsilon) \gets value$
\State $(P_1,\ldots,P_{n-1}) \gets o_u$;  $H \gets \hash_k(pwd_g,s_u)$
\For{$m=1,\ldots,n-1$}
\State {\bf If $P_m(H)=1$ then Break}
\State $H \gets \hash_k(H)$
\EndFor
\State {\bf SendToServer}$(u,H)$
\end{algorithmic}
\caption{Reproduce (Client Side)}
\label{alg:Reproduce}
\end{algorithm}

Here, we use $\stopcond(pwd,o)$ to denote the implicitly defined stopping time for each password $pwd \in \pwdspace$ given the sequence $o=\big(\upred{1},\ldots, \upred{n-1}\big)$ of halting predicates. Formally,  $\stopcond(pwd,o) = i$ if and only if (1) $\upred{i}\left(\hash_k^{i}\left(pwd,s_u \right) \right)=1$ and $\upred{j}\left(\hash_k^{j}\left(pwd,s_u \right) \right)=0$ for all $j < i$ or (2) $i=n$ and $\upred{j}\left(\hash_k^{j}\left(pwd,s_u \right) \right)=0$ for all $j < n$, where $n$ denotes the maximum possible rounds of hash iteration. 

Thus, to compute $\hash_k^{\stopcond(pwd_g,o_u)}(pwd_g,s_u)$ we initially compute $H_1 \gets \hash_k(pwd_g,s_u)$ and check if $\upred{1}(H_1)=1$. If it is then we return the derived password $H_1$. Otherwise, we compute $H_{2} \gets \hash_k(H_1)$ and return $H_2$ if and only if $\upred{2}(H_1)=1$. This process is repeated until either  $\upred{i}(H_i)=1$ or $i=n$. The derived password is sent to the server to be accepted or rejected. Authentication is guaranteed when $pwd_g = pwd_u$ and is very unlikely when $pwd_g \neq pwd_u$ because $\hash$ is collision resistant. The client-side algorithm \reproduce is presented formally as Algorithm \ref{alg:Reproduce}. We note that, unlike the account creation process, the algorithm \reproduce  is entirely deterministic. 

\noindent{\em Remark. } We omit any description of how the authentication server stores the derived password as this is an orthogonal issue. In an ideal world the authentication server would add salt and apply a strong key-stretching algorithm before storing the derived password. Unfortunately, many authentication servers have failed to adopt these standard security practices~\cite{bonneau2010password}. Furthermore, users will not necessarily know what security practices have been adopted until the authentication server is actually breached. Thus, in our security analysis we will assume that the authentication server does not do any key-stretching. By applying salting and key-stretching algorithms on the client-side we can help protect users even when organizations fail to adopt these security practices. Of course if the authentication server does perform additional key stretching then the adversary's task will be even harder.

\noindent{\bf Notation and Customizable Parameters} We use $B$ to denote the budget of the adversary (i.e., the maximum number of times the adversary is willing to evaluate $\hash$ in an attempt to crack the user's password), and we will use $C_{srv}$ to denote the maximum cost that the client is willing to bear per authentication session (in expectation).

\cut{We use $n$ to denote the maximum number of rounds of hashing. Specifically, Client-CASH will use $t$ iterations of $\hash_k$ for some $t \in \{1, \ldots,n\}$. We will use halting predicates to determine when to stop iterating $\hash_k$.  Here, a predicate $\upred{}:\{0,1\}^*\rightarrow\{0,1\}$ is simply a function that tells us when to stop iterating $\hash_k$, and the predicate sequence $\upred{1},\ldots,\upred{n-1}$ is stored on the client along with a random $L$ bit salt value $s_u \gets \mathbf{Unif}\left(\{0,1\}^L\right)$. Here $\mathbf{Unif}(S)$ denotes a randomized algorithm that samples from the set $S$ uniformly at random.  }

We use  $\outputspace$ to denote the range of the $\selpred$ function (e.g., the space of all valid predicate sequences of length $n-1$, where $n$ denotes the maximum number of rounds of hashing for $\reproduce$). 
 We adopt the following notational conventions: Given a randomized algorithm like $ \selpred$ we use $o \leftarrow \selpred(pwd)$ to denote a random sample from the distribution induced by an input $pwd$. If we fix the random input bits $r$ then we will use $o:= \selpred(x; r)$ to denote the deterministic result. We will use $\pwdspace$ to denote the space of all possible passwords that the user might select.  

For a user with password $pwd_u$ and predicate sequence $o_u$, the hash used by that user on the instance of a correct password is $\hash_k^{\stopcond(pwd_u,o_u)}$. We will use $\outputspace_{j,pwd}$ $\doteq$ $\left\{o \in \outputspace~\vline~\right.$ $\left.\stopcond(pwd,o) =j\right\} \subseteq \outputspace$ to denote the subset of outcomes which yield stopping time $j$ for the password $pwd$.  Finally, we will use the parameter $\epsilon > 0$ to quantify the maximum amount of information leaked about a user's password by the output $o \in \outputspace$ of the  $\selpred$ function. For readers familiar with the notion of $\epsilon$-differential privacy we remark that we use the same notation intentionally.  

\cut{
\begin{table}[t]
\centering
\begin{tabular}{|c|p{4in}|}
\hline
Term & Description \\
\hline
$\hash$ & Cryptographic hash function\\
\hline
$\hash^t$ & Hash function with $t$ iterations\\
\hline
$k$ & The number of hash iterations in each round of \clientcash\\
\hline
$n$ & The maximum number of hashing rounds. $nk$ is the maximum number of hash iterations\\
\hline
$a$ & Account name\\
\hline
$C_{srv}$ & Maximum amortized the cost the client is willing to bear\\
\hline
$C_H$ & Computational cost of one iteration of the hash $\hash$\\
\hline
$\pwdspace$ & Space of all possible passwords\\
\hline
$\outputspace$ & The set of all valid predicate sequences\\
\hline
$\outputspace_{j,pwd}$ & The set of outcomes with stopping time $j$ given a fixed $pwd \in \pwdspace$.\\
\hline
$O_j$ & The size of $\outputspace_{j,pwd}$ for symmetric $\outputspace$. \\
\hline
\cut{$\prob{o}$ & The probability that $o \in \outputspace$ is selected by \selpred for a fixed $pwd$. \\
\hline}
$\prob{j}$ & The probability of each outcome $o \in \outputspace_{j,pwd}$.\\
\hline
$\hprob{j}$ & The probability that a given password $pwd$ has stopping time $j$.\\
\hline
$s_u$ & The random salt for user $u$. \\
\hline
$\stopcond(pwd,o)$ & The stopping time of $pwd$ given $o \in \outputspace$. The derived hash of the password is $\hash^{\stopcond(pwd,o)\times k}(pwd,s)$. \\
\hline
$s \stackrel{\$}{\gets}\{0,1\}^L$ & The salt $s$ is randomly selected uniformly from $\{0,1\}^L$. \\
\hline
\end{tabular}
\caption{Notation}
\label{tab:Notation}
\end{table}}

%\section{System Construction}
%\label{sec:System Overview}
%\input{system_overview}

%\vspace{-0.4cm}
\section{Design of the Mechanism}
%\vspace{-0.4cm}
\label{sec:sys_properties}
% !TEX root = ASIA_CCS_PWD_SYSTEM_2016.tex
In the previous section we outlined the \clientcash~mechanism using the randomized algorithm \selpred~in a black-box manner. We now examine the 
exact formulation of this algorithm.

%\vspace{-0.4cm}
\subsection{Security Requirements}
%\vspace{-0.2cm}
The probability that $\selpred{(pwd)}$ yields a particular outcome $o \in \outputspace$ may depend on the input password $pwd$. Indeed, our goal is to introduce a cost asymmetry so that, in expectation, the stopping time $\stopcond(pwd,o)$ for the correct password is less than stopping time $\stopcond(pwd',o)$ for any incorrect passwords $pwd'\neq pwd$. One natural way to achieve this asymmetry might be to define a family of predicates $$P_{pwd,j}(x)= 
\begin{cases}
1 &\text{if } x \equiv H_k^j\left(pwd,s_u \right) \\
0 &\text{else}
\end{cases}
$$
for each password $pwd \in \pwdspace$ and $j < n$. Then $\selpred(pwd)$ might select a stopping time $j$ at random from some distribution and set $o = \upred{1},\ldots,\upred{n-1}$ where $\upred{i} = P_{pwd,j}$ for each $i <n$. This solution could provide an extreme cost asymmetry. In particular, we would have $\stopcond(pwd,o) = j$ for the correct password and with high probability we would have $\stopcond(pwd',o) = n$ for all other passwords. Indeed, this solution is simply a reformulation of Boyen's~\cite{boyen2007halting} halting puzzles.  

The downside to this solution is that it would enable an adversary to mount an offline attack using {\em only} state from the client (e.g., without breaching the authentication server) because the output $o$ of $\selpred$ is stored on the client and this output would include the hash value $ H_k^j\left(pwd,s_u \right)$. 

In contrast to Boyen~\cite{boyen2007halting}, we will require the $\selpred$ function to satisfy a very stringent information theoretic security property. In particular, for every $o \in \outputspace$ and for all $pwd,pwd' \in \pwdspace$ with $pwd \ne pwd'$, we will require that \begin{equation} \label{eq:Security}
\frac{\Pr\left[\selpred{\left(pwd\right)}=o\right]}{\Pr\left[\selpred{\left(pwd'\right)}{}=o\right] } \leq e^\epsilon \ .
\end{equation}
Here, $\epsilon$ is the security parameter which upper bounds the amount of information about the user's password that is leaked by the outcome $o$.  This requirement is exactly equivalent to the powerful notion of $\epsilon$-differential privacy~\cite{dwork2006calibrating}. Intuitively, the probability that we produce a particular outcome $o \in \outputspace$ cannot depend too much on particular password $pwd$ that the user chose. 

{\bf Discussion.} Differential privacy is an information theoretic guarantee, which holds even if the adversary has background knowledge about the user. The security guarantees are quite strong. In particular, let $\mathbf{Attack}_z$ denote {\em any} attack that an adversary with background knowledge $z$ might use after observing the outcome $o$ from $\selpred(pwd)$ and let $\mathbf{Bad} \subseteq \mathbf{Range}\left(\mathbf{Attack}_z \right)$ denote the set of outcomes that our user Alice would consider harmful. It is easy (e.g., see \cite{dwork2013algorithmic}) to show that 
$$
\frac{\Pr\left[\mathbf{Attack}_z\left(\selpred{\left(pwd\right)}\right)\in \mathbf{Bad}\right]}{\Pr\left[\mathbf{Attack}_z\left(\selpred{\left(pwd'\right)}\right)\in \mathbf{Bad}\right]} \leq e^\epsilon  \ .
$$
Intuitively, the numerator represents the adversary's actual success rate and the denominator represents the adversary's success rate if the output of $\selpred$ were not even correlated with Alice's real password. Thus, the constraint implies that the probability the adversary's attack succeeds cannot greatly depend on the output of the $\selpred$ function. 

%\vspace{-0.4cm}
\subsection{Cost Requirements}
%\vspace{-0.15cm}
Performing client-side key stretching requires computational resources on the client side. We will require that the amortized cost of hashing a password does not exceed the maximum amortized cost the user is willing to bear. Let $C_{srv}$ denote the user's maximum amortized cost and let $k\cdot C_H$ denote the cost of executing $\hash_k$ one time. Finally, for a fixed password $pwd$ let $\hprob{j,pwd}\doteq  \Pr\left[\selpred(pwd) \in \outputspace_{j,pwd}  \right]$ denote the probability that $\selpred$ returns an outcome $o \in\outputspace$ which yields stopping time  $\stopcond(pwd,o)=j$ for the password $pwd$. Given a password $pwd$ the expected cost of executing $\reproduce$ in \clientcash~ is 
$$\mathbf{E}\left[ \stopcond(pwd,o)\cdot k \cdot C_H \right] = k\cdot C_H \sum\limits_{i=1}^n  i \cdot \hprob{i,pwd} \ ,$$ where the expectation is taken over the randomness of the algorithm $\selpred$.  Thus, we require that 
 \begin{equation} \label{eq:CostConstraint}
\sum\limits_{i=1}^{n} i\cdot\hprob{i,pwd} \le \alphak  \ .
\end{equation}

%\vspace{-0.4cm}
\subsection{Symmetric Predicate Sets}
%\vspace{-0.2cm}
In this section we show how to simplify the security and cost constraints from the previous sections by designing $\outputspace$ with certain symmetric properties. In particular, we need to design the $\selpred$ algorithm in such a way that the cost constraint (eq. \ref{eq:CostConstraint}) is satisfied for 
{\em all passwords} $pwd \in \pwdspace$, and we need to ensure that our security constraint (eq. \ref{eq:Security}) holds for {\em all pairs} of passwords $pwd \ne pwd'$. Working with $\left|\pwdspace \right|^2$ different security constraints is unwieldy. Thus, we need a way to simplify these requirements. 

 One natural way to simplify these requirements is to construct a \textit{symmetric} $\outputspace$; that is, for all $pwd,pwd' \in \pwdspace$ and for all $j \in \{1,\ldots,n\}$, $|\outputspace_{j,pwd}| = |\outputspace_{j,pwd'}|$. We stress that this restriction does not imply that the sets $\outputspace_{j,pwd}$ and $\outputspace_{j,pwd'}$ are the same, simply that their sizes are the same. We will also require that $ \forall pwd,pwd' \in \pwdspace, \forall j \leq n, \forall o \in O_{j,pwd}, o' \in O_{j,pwd'} $
\begin{eqnarray} \label{eq:symmetry}
\Pr[\selpred(pwd) = o] = \Pr[\selpred(pwd') = o'] \end{eqnarray}  as there is no clear reason to favor one outcome over the other since both outcomes $o,o' \in \outputspace$ yield the same halting time for their respective passwords $pwd$ and $pwd'$ (i.e, $\stopcond(pwd,o) = \stopcond(pwd',o')$) . 

\noindent{\bf Notation for Symmetric $\outputspace$.} Given a symmetric $\outputspace$, we define $O_j \doteq |\outputspace_{j,pwd}|$. We define $\prob{j} \doteq \Pr[\selpred(pwd) = o]$ for an arbitrary password $pwd \in \pwdspace$ and output $o \in \outputspace_{j,pwd}$ since, by symmetry, the choice of $pwd$ and $o$ do not matter. Intuitively, $\prob{j}$ denotes the probability that $\selpred(pwd)$ outputs $o$ with stopping time $\stopcond(pwd,o) = j$. We will also write $\hprob{j}$ instead of $\hprob{j,pwd}=O_j\cdot \prob{j}$ because the value will be the same for all passwords $pwd \in \pwdspace$.

\cut{We can similarly write the rest of the system properties in terms of the $O_j$'s and $\prob{j}$'s which gives us a more convenient way of representing the mechanisms independent of the specific password used. A full list of properties and descriptions using both the traditional and equivalence class notations are shown in Table 2. In the \selpred algorithm, the probability distribution $\mathscr{D}$ over the elements of $\outputspace$ is thus dependent on $(\prob{1},\ldots,\prob{n})$ (which is accessible to the adversary) as well as the partition$\{\outputspace_{j,pwd_u}\}$ which can only be computed when $pwd_u$ is known. \\}

Now the $\selpred$ algorithm can be completely specified by the values $\prob{j}$ for $j \in \{1,\ldots,n\}$. It remains to construct a symmetric $\outputspace$. The construction of a symmetric $\outputspace$ used in this work is simple we will focus on stopping predicates with the following form: 
$$
P_{i,j}(x) = 
\begin{cases}
1 &\text{if } x \equiv i \mod j\\
0 &\text{else}
\end{cases}
$$

Theorem \ref{thm:symmetric_output} says that we can use these predicates to construct a symmetric set $\outputspace$. 

\begin{theorem}
\label{thm:symmetric_output}
Let integers $\ell_{1},\ldots,\ell_{n-1} \geq 2$ be given and suppose that $\outputspace = \{P_{0,\ell_{1}},\ldots P_{\ell_1-1,\ell_{1}}\} \times$ $\ldots$ $\times$ $\{P_{0,\ell_{n-1}}$,$\ldots$, $P_{\ell_{n-1}-1,\ell_{n-1}}\}$. Then  $\outputspace$ is symmetric, i.e. for all $j \in \{1,\ldots,n\}$ and for all $pwd,pwd' \in \pwdspace$ we have $|\outputspace_{j,pwd}| = |\outputspace_{j,pwd'}|$. 
\end{theorem}

One of the primary advantages of using a symmetric $\outputspace$ is that we can greatly reduce the number of security and cost constraints. Table \ref{tab:Constraints} compares the security and cost constraints with and without symmetry. Without symmetry we had to satisfy separate cost constraints for all passwords $pwd \in \pwdspace$. With symmetry we only have to satisfy one cost constraint.  Similarly, with symmetry we only need to satisfy one security constraint for each $i,j \in \{1,\ldots,n\}$ instead of multiple constraints for each pair of passwords $pwd \ne pwd'$ that the user might select. While the space of passwords may be very large, $n$, the number of rounds of hashing, will typically be quite small (e.g., in this paper $n \in \{2,3\}$).  \fullversion{See Theorems \ref{thm:diff_privacy} and \ref{thm:cost_constraints} in the appendix for formal statements of these results and their proofs.} \confversion{See the full version of this paper for proofs of the statements in Table \ref{tab:Constraints}.}

In our analysis of \clientcash~we will focus on two simple symmetric constructions of $\outputspace$:
\begin{itemize}
\item ``Two Round" Case: set $n=\ell_1=2$ in Theorem \ref{thm:symmetric_output} so $\outputspace =  \{P_{0,2},P_{1,2}\}$.
\item ``Three Round" Case: set $n=\ell_1=\ell_2 = 3$ in Theorem \ref{thm:symmetric_output} so $\outputspace = \{P_{0,3},P_{1,3},P_{2,3}\}^2$. 
\end{itemize}
We do not rule out the possibility that other cases would yield even stronger results. However, we are already able to obtain significant reductions in the adversary's success rate with these simple cases.

\cut{
\newcommand{\thmdiffprivacy}{
Suppose that $\outputspace$ is constructed as specified in Theorem \ref{thm:symmetric_output}, equation \ref{eq:symmetry} holds and $\forall i,j \in \{1,\ldots,n\}, \frac{\prob{i}}{\prob{j}} \le e^\epsilon$. Then $\forall pwd,pwd' \in \pwdspace, \forall o \in \outputspace, \frac{ \Pr[\selpred(pwd) = o]}{\Pr[\selpred(pwd') = o]} \le e^\epsilon$. 
}

\begin{theorem}
\label{thm:diff_privacy}
\thmdiffprivacy
\end{theorem}}

\cut{
\newcommand{\thmcostconstraints}{
Suppose that $\outputspace$ and the $\prob{i}$'s are constructed as specified in Theorems \ref{thm:symmetric_output} and \ref{thm:diff_privacy}, and that $\sum\limits_{i=1}^n i \cdot O_i \prob{i} \le \alphak$. Then for all $pwd \in \pwdspace$ we have $\sum\limits_{i=1}^n i \cdot \hprob{i,pwd}  \le \alphak$. 
}

\begin{theorem}
\label{thm:cost_constraints}
\thmcostconstraints
\end{theorem}}

\cut{
From a security perspective, this method of finding the distribution used to select $o_u \in \outputspace$ depends on the user's choice of password $pwd_u$. Without this password, the adversary can only compute $\hprob{j}$'s and not the probabilities of selecting individual $o \in \outputspace$. }

\begin{table}[t]
\centering
\begin{tabular}{|c|c|c|}
\hline
No Symmetry & Constraint & With Symmetry \\
\hline
\cut{$\forall o \in \outputspace, \prob{o} \in [0,1]$ & probability constraint & $\forall i \in \{1,\ldots,n\}, \prob{i} \in [0,1]$\\
$o,o' \in \outputspace_{j,pwd} \Rightarrow \prob{o} = \prob{o'}$ & &\\
& & \\
$\sum\limits_{o \in \outputspace} \prob{o} = 1$ & probability constraint & $\sum\limits_{i=1}^n O_i \prob{i} = 1$\\
&&\\}
$\forall pwd,pwd' \in \pwdspace,\forall o \in \outputspace, $ & Security  &  $\forall i,j \in \{1,\ldots,n\},$ \\
$ \frac{\Pr[\selpred(pwd)=o]}{\Pr[\selpred(pwd') = o]} \le e^\epsilon$ &  \fullversion{(Thm  \ref{thm:diff_privacy})}\confversion{\cite[Thm 4]{fullVersion}}   & $\frac{\prob{i}}{\prob{j}} \le e^{\epsilon}$\\
&  &\\
\hline
\cut{$\forall j \in \{1,2,...n\}, \hprob{j} = \Pr[\selpred(pwd)=\outputspace_{j,pwd}]  $ & Stopping time probabilities & $\forall j \in \{1,\ldots,n\}, O_j\prob{j} = \hprob{j}$\\
& &\\}
 $\forall pwd \in \pwdspace,$  & & \\
 $\sum\limits_{j=1}^n j\cdot \hprob{j,pwd} \le \alphak$ & Cost & $\sum\limits_{j=1}^n j\cdot O_j\prob{j} \le \alphak$\\
 & \fullversion{(Thm \ref{thm:cost_constraints})}\confversion{\cite[Thm 5]{fullVersion}} & \\
 \hline
\end{tabular}
\vspace{-0.2cm}
\caption{Security and cost constraints for $\selpred$ with symmetric predicate set $\outputspace$.}
\label{tab:Constraints}

\end{table}

\cut{
\begin{algorithm}[tbh]
\begin{algorithmic}
\State {\bf Input:} $pwd_u;\text{random bits } r$
State {\bf Parameters: } $\outputspace, n$
\State $\{\outputspace_{j,pwd_u}\}_{j \in \{1,\ldots,n\}} \gets \mathbf{ComputeStoppingTimes}(pwd_u,n)$
\State $S \gets \{\outputspace_{j,pwd_u}\}_{j \in \{1,\ldots,n\}}$
\State $o_u \gets \mathscr{D}(S,r)$
\State {\bf Return} $o_u$
\end{algorithmic}
\caption{SelectPredicates (Client Side) \jnote{Where is the $\mathscr{D}$ notation defined? Possibly move to appendix}}
\label{alg:SelectPredicates}
\end{algorithm} 
}

%\vspace{-0.4cm}
\section{The Adversary Model}
%\vspace{-0.4cm}
\label{sec:The Adversary}
% !TEX root = ASIA_CCS_PWD_SYSTEM_2016.tex
In this section we formalize the strategies that an offline adversary might adopt so that we can analyze the effectiveness of the \clientcash~mechanism against an offline adversary. In an \textit{online} attack, the adversary must interact with and query the authentication server to have a password verified or rejected, and there are a variety of effective methods to deal with such an attack (e.g., a $k$-strikes policy). By contrast, the offline adversary we consider is limited only by the resources that he is willing to invest to crack the user's password. The adversary's goal is to maximize the odds that he cracks the user's password given a finite computational budget $B$. We will use $\padvB$ (resp. $\pdetB$) to denote the probability that the {\em optimal} offline adversary successfully cracks a password protected with \clientcash~(resp. deterministic key-stretching) with a budget $B$ denoting the maximum number of times that the adversary is willing to compute $\hash$. To simplify notation we will often write $\padv$ (resp. $\pdet$) when $B$ is clear from context. 

\noindent {\em The Adversary's Knowledge.} We consider an adversary who has breached the client and one of the third party authentication servers (e.g., LinkedIn) where the user has an account. This adversary has access to all code and data on the client and on a third party authentication server. Suppose for example that user $u$ created account $a$ with a password $pwd_u$. The client will store the tuple $(u,a,o_u,s_u,n,\outputspace,\epsilon)$ and the third party authentication server will have some record of the derived password $\hash_k^{\stopcond\left(pwd_u,o_u\right)}\left(pwd_u,s_u \right)$. In our analysis we do not assume that the third party authentication server adopts proper password storage procedures like key-stretching --- this is the unfortunate reality for many authentication servers~\cite{bonneau2010password}. If the authentication server does apply key-stretching then the adversary's offline attack will be even more expensive. 

\paragraph{Offline attack against the traditional mechanism} We first analyze the adversary's success rate $\pdetB$ against a traditional deterministic key stretching algorithm with comparable cost to \clientcash. In particular, let $k' = \frac{C_{srv}}{C_H}$. Instead of adopting \clientcash~the adversary might instead use a deterministic key-stretching algorithm $\hash_{k'}$ with the same cost. In this case the adversary could check at most $\frac{B}{k'}$ passwords, and each password has probability $\frac{1}{|\pwdspace|}$ of being the correct password.  Thus, \begin{equation} \label{eq:pdet}
\pdetB = 
\begin{cases}
\frac{B}{k'|\pwdspace|} & \text{if } \frac{B}{k'} \le |\pwdspace|\\
1 & \text{otherwise}
\end{cases}
\end{equation}

\paragraph{Offline attack against \clientcash} We now consider the adversary's optimal success rate $\padvB$ against \clientcash. While each password $pwd \in \pwdspace$ is chosen with equal probability, we might have different stopping times $\stopcond\left(pwd,o_u\right)$ for each password. The adversary can compute $\hash_k$ up to $\frac{B}{k}$ times so he can complete $B/k$ rounds of hashing in total. For each password guess $pwd_g$ the adversary can decide how many rounds of hashing to complete for that particular guess. There are many different strategies that the adversary might adopt, and some may produce a larger probability of success than others.  Of course the optimal adversary will never complete more rounds of hashing than necessary (e.g., if $\stopcond(pwd_g,o) = i$ then the adversary will never complete $i+1$ rounds of guessing for guess $pwd_g$). 

It turns out that we can represent the optimal adversary's strategies as a vector of $n$ numbers $\mathbf{b} = (b_1,\ldots,b_n) \in \mathbb{R}^n$ where $b_i \in \left[0, \min\left\{1,k\left|\pwdspace\right|/B \right\} \right]$ denotes the fraction of the adversary's budget spent hashing passwords $i$'th round \fullversion{(see the appendix for a more formal justification)}\confversion{(see the full version of this paper for a more formal justification)}. Formally, given a symmetric set $\outputspace$ from Theorem \ref{thm:symmetric_output} we use $F_B \subseteq \mathbb{R}^n$ to denote the set of feasible strategies for the adversary 
\confversion{\begin{eqnarray*}
F_B &=& \left\{\mathbf{b} \in \mathbb{R}^n~\vline  ~\sum_{i=1}^n b_i = 1 \bigwedge \forall i\leq n  \right. \\
& & ~~\left. \left( b_i \in \left[ 0, \min\left\{1,\frac{k\left|\pwdspace\right|}{B}  \right\} \right]   \bigwedge b_{i+1} \leq b_i - \frac{b_i}{\ell_i}\right) \right\} \ .
\end{eqnarray*}}\fullversion{\begin{eqnarray*}
F_B &=& \left\{\mathbf{b} \in \mathbb{R}^n~\vline  ~\sum_{i=1}^n b_i = 1 \bigwedge \forall i\leq n  \right.  ~~\left. \left( b_i \in \left[ 0, \min\left\{1,\frac{k\left|\pwdspace\right|}{B}  \right\} \right]   \bigwedge b_{i+1} \leq b_i - \frac{b_i}{\ell_i}\right) \right\} \ .
\end{eqnarray*}}
We require that $b_{i+1} \leq b_i\cdot \frac{\ell_{i}-1}{\ell_i}$ because the adversary can only hash a password $pwd_g$ in round $i+1$ if he previously completed round $i$ for the same password and the optimal adversary will not hash $pwd_g$ in round $i+1$ if $\stopcond(pwd_g,o)=i$ --- for a random password the halting predicate will output $1$ with probability $1/\ell_i$ in the $i$'th round. Given a particular strategy $\mathbf{b}$ the adversary succeeds with probability $$\pwdratio \left( \hprob{1}b_1 + \sum\limits_{i=2}^{n} \hprob{i}b_i \prod\limits_{j=1}^{i-1}\left( \frac{\ell_j}{\ell_j-1} \right) \right) \ .$$ 
Inutuitively, $b_i \pwdratio$ denotes the fraction of password selected by the adversary to be hashed at least $i$ time, and the $\prod\limits_{j=1}^{i-1} \left( \frac{\ell_j-1}{\ell_j}\right)$ term represents the fraction of passwords with stopping time $\stopcond(pwd_g,o)>i-1$. Thus, $\hprob{i}\prod\limits_{j=1}^{i-1} \left( \frac{\ell_j}{\ell_j-1}\right)$ denotes the conditional probability that $\stopcond(pwd_g,o)=i$ given that $\stopcond(pwd_g,o)>i-1$.

Because the user selects passwords uniformly at random from the set $\pwdspace$ we might conjecture that the optimal adversary will follow the same strategy for every password guess (e.g.,  for some $i\leq n$ the optimal adversary will hash each guess $pwd_g$ for $\min\{i,\stopcond(pwd_g,o)\}$ rounds before moving on to the next guess). This intuition turns out to be correct. Formally, $\advstrat{i} \doteq \{\mathbf{b} \in F_B: \forall \mathbf{b}' \in F_B, b_i \ge b_i'\}$ denotes the set of strategies in which the adversary tries each password guess for $\min\{i,\stopcond(pwd_g,o)\}$ rounds before giving up.  Our next results states that for some $i \leq n$ the adversary will follow a strategy in $\advstrat{i}$.

\newcommand{\thmfullpadv}{
Assume that $\outputspace$ is constructed as in Theorem \ref{thm:symmetric_output}. The dominant adversary strategies are given by the collection of sets $\advstrat{i} = \{\mathbf{b} \in F_B: \forall \vec{b}' \in F_B, b_i \ge b_i'\}$ for $1 \le i \le n$. Let $F_B^* := \bigcup\limits_{i=1}^n \advstrat{i}$. Then $$
\padvB = \pwdratio \max_{\mathbf{b} \in F_B^*} \left\{  \hprob{1}b_1 + \sum\limits_{i=2}^n \hprob{i}b_i \prod\limits_{j=1}^{i-1} \left( \frac{\ell_j}{\ell_j-1}  \right)   \right\}
$$
}

\begin{theorem}
\label{thm:fullpadv}
\thmfullpadv
\end{theorem}
\noindent{\em Discussion.} In our analysis of the adversary we assume that the user picks passwords uniformly at random from $\pwdspace$. Thus, we can only recommend \clientcash~to users for whom this assumption holds. A large body of research has explored the security of user selected passwords (e.g., \cite{massey1994guessing,boztas1999entropies,pliam2000incomparability,bonneau2012science}), how users cope with multiple passwords \cite{florencio2007large} and how users respond to password restrictions \cite{usability:compositionPolicies,blockiPasswordComposition}. These results indicate that many users do not select their password uniformly at random --- contrary to the assumption we made in our security analysis. However, several research results indicate that users are capable of remembering truly random system-assigned passwords with practice \cite{blockiNaturallyRehearsingPasswords,spacedRepetitionAndMnemonics,BS14,usabilitystudy:xkcd}. Our assumption would hold for users who adopt the four random words strategy popularized by the web comic XKCD\footnote{See \url{http://xkcd.com/936/} (retrieved 9/8/2015).} or the password management strategies proposed by Blocki et al.~\cite{blockiNaturallyRehearsingPasswords}. Thus, we believe that this assumption could be reasonable for many the security conscious users who would adopt \clientcash.

\subsection{Two Round Strategies}
% !TEX root = ASIA_CCS_PWD_SYSTEM_2016.tex
In the two-round case we have $F^* = \advstrat{1} \cup \advstrat{2}$ in  Theorem \ref{thm:fullpadv}, so it suffices to calculate these two sets. We have $\advstrat{2} = \left\{ \left( \frac{2}{3},\frac{1}{3} \right) \right\}$ and 
 $$
\advstrat{1} = 
\begin{cases}
\{(1,0)\} & \text{if } \pwdratio \le 1\\
\left\{\left( \invpwdratio, 1- \invpwdratio\right)\right\} & \text{otherwise.}
\end{cases}
$$
If $\mathbf{b} = (1,0)$ then, using the fact that $O_1 = 1$, we have \[\pwdratio\left( 
O_1\prob{1} b_1 + \sum\limits_{i=2}^n O_i\prob{i}b_i \prod\limits_{j=1}^{i-1} \left(\frac{\ell_j}{\ell_j - 1} \right)\right) = \prob{1}\pwdratio \ .\] 
 Thus, working out each case for $\mathbf{b} \in F^*$  Theorem \ref{thm:fullpadv} simplifies to: $$
\padvB \ge 
\begin{cases}
\max \left\{\prob{1}\pwdratio, \frac{2}{3}\pwdratio \right\} &\text{if } \pwdratio \le 1\\
\max \left\{ \prob{1} + 2\left(\pwdratio - 1\right)\prob{2}, \frac{2}{3}\pwdratio \right\} & \text{otherwise. }
\end{cases}
$$

\subsection{Three Round Strategies}
% !TEX root = ASIA_CCS_PWD_SYSTEM_2016.tex

%Where, in the last equation, the inequality is changed to equality as discussed in \textsection \ref{sec:AnalysisoftheExponentialMechanism}. 
In the three-round case we have $F^* = \advstrat{1} \cup \advstrat{2} \cup \advstrat{3}$ in  Theorem \ref{thm:fullpadv}, so it suffices to calculate these three sets. To bound $\padvB$ we note that it is sufficient to find the extremal points of each region since the adversary's objective function is linear in the $b_i$'s. Letting $\overline{\advstrat{i}}$ denote the extremal points of $\advstrat{i}$, we thus have 

\begin{equation*}
\begin{split}
\overline{\advstrat{1}} =&
\begin{cases}
\{(1,0,0)\} & \text{if } \pwdratio \le 1\\
\left\{  \left( \invpwdratio,1-\invpwdratio,0\right),  \right. & \\
\left.  \left(\invpwdratio,\frac{3}{5}\left(1-\invpwdratio\right),  \frac{2}{5} \left(1-\invpwdratio\right)    \right)     \right\} &\text{if } \pwdratio \in \left[1,\frac{5}{3}\right]\\
\left\{   \left(  \invpwdratio,\frac{3}{5}\left(1-\invpwdratio\right),\frac{2}{5} \left(1-\invpwdratio\right)    \right)     \right\} & \text{otherwise}
\end{cases}\\
\end{split}
\end{equation*}
\begin{equation*}
\begin{split}
\overline{\advstrat{2}} = &
\begin{cases}
\left\{   \left(\frac{3}{5},\frac{2}{5},0\right)  \right\} &\text{if } \pwdratio \le \frac{5}{3}\\
\left\{  \left( \invpwdratio,\frac{2}{3}\invpwdratio,1-\frac{5}{3}\invpwdratio   \right)  \right\} &\text{otherwise}
\end{cases}\\
\overline{\advstrat{3}} = &\left\{  \left(\frac{9}{19},\frac{6}{19},\frac{4}{19} \right) \right\}
\end{split}
\end{equation*}

Thus, we can express $\padvB$ in terms of the following bounds. \\
\confversion{\begin{align*} 
&(1)~3\prob{1}\pwdratio ~~
&(2)&~\left( \frac{9}{5}\prob{1} + \frac{6}{5}\prob{2}  \right)\pwdratio\\
&(3)~\frac{9}{19}\pwdratio 
&(4)&~3\prob{1} + 3\prob{2}\left( \pwdratio - 1\right) 
\end{align*}
\begin{align*} 
&(5) ~3\prob{1} + \left( \frac{9}{5}\prob{2} + \frac{18}{5}\prob{3}  \right)\left(\pwdratio - 1\right) ~~~~~~~~~~~\\
&(6)~ 3\prob{1} + 2\prob{2} + 9\prob{3}\left( \pwdratio - \frac{5}{3}  \right)  
\end{align*}}
\fullversion{
\begin{align*} 
&(1)~3\prob{1}\pwdratio ~~
&(2)&~\left( \frac{9}{5}\prob{1} + \frac{6}{5}\prob{2}  \right)\pwdratio
&(3)&~\frac{9}{19}\pwdratio  \\
&(4)~3\prob{1} + 3\prob{2}\left( \pwdratio - 1\right) 
&(5)& ~3\prob{1} + \left( \frac{9}{5}\prob{2} + \frac{18}{5}\prob{3}  \right)\left(\pwdratio - 1\right) 
&(6)&~ 3\prob{1} + 2\prob{2} + 9\prob{3}\left( \pwdratio - \frac{5}{3}  \right)  
\end{align*}
}
For example,  if $\mathbf{b}=\left(\frac{3}{5},\frac{2}{5},0\right) $ then, using the fact that $O_1 = 3, O_2=2$ and $\ell_j = 3$, we have 
\confversion{\begin{eqnarray*}
\pwdratio \left(O_1\prob{1} b_1 + \sum\limits_{i=2}^n O_i\prob{i}b_i \prod\limits_{j=1}^{i-1} \left(\frac{\ell_j}{\ell_j - 1} \right)\right) 
 \end{eqnarray*}
\begin{eqnarray*}
 &=&  \pwdratio\left(3\prob{1} \frac{3}{5} +  2\prob{2}\frac{2}{5}  \left(\frac{3}{2} \right) \right)  =  \pwdratio\left(\prob{1} \frac{9}{5} +  \frac{6}{5}\prob{2} \right)  \\
\end{eqnarray*}}\fullversion{\begin{eqnarray*} \pwdratio \left(O_1\prob{1} b_1 + \sum\limits_{i=2}^n O_i\prob{i}b_i \prod\limits_{j=1}^{i-1} \left(\frac{\ell_j}{\ell_j - 1} \right)\right) 
 &=&  \pwdratio\left(3\prob{1} \frac{3}{5} +  2\prob{2}\frac{2}{5}  \left(\frac{3}{2} \right) \right)  =  \pwdratio\left(\prob{1} \frac{9}{5} +  \frac{6}{5}\prob{2} \right)  \\
\end{eqnarray*}}
as in (2). Thus, in the three-round case Theorem \ref{thm:fullpadv} simplifies to:
$$
\padvB \ge 
\begin{cases}
\max\{(1),(2),(3)\} &\text{if } \pwdratio \le 1\\
\max\{ (2),(3),(4),(5)   \} &\text{if } \pwdratio \in \left[1,\frac{5}{3}\right]\\
\max\{  (3),(5),(6)   \} &\text{otherwise.  }
\end{cases}
$$

\newcommand{\thmpadv}{Assuming that $\outputspace$ is constructed as in in Theorem \ref{thm:symmetric_output}
$$
\padvB = \pwdratio \max\limits_{\mathbf{b} \in F  }\left\{ \hprob{1}b_1 + \sum\limits_{i=2}^{n} \hprob{i}b_i \prod\limits_{j=1}^{i-1}\left( \frac{\ell_j}{\ell_j-1} \right)   \right\}
$$
}

\newcommand{\corpadv}{
When $\ell_i = n$ for all $i \in \{1,\ldots,n\}$, $$
\padvB = \pwdratio\max\limits_{ \mathbf{b} \in F }\left\{ \sum\limits_{i=1}^n \hprob{i}b_i \left(  \frac{n}{n-1} \right)^{i-1}  \right\}
$$
}

%\subsection{Optimal Adversary Strategies}
%\label{subsec:Optimal Strategies}
%\input{adv_opt_strat}
%\vspace{-0.4cm}
\section{The Exponential Mechanism}
%\vspace{-0.4cm}
\label{sec:AnalysisoftheExponentialMechanism}

% !TEX root = ASIA_CCS_PWD_SYSTEM_2016.tex
In this section we show how to construct the $\selpred$ function so that it satisfies our security and cost constraints (equations \ref{eq:Security} and \ref{eq:CostConstraint}). Our construction is based on the Exponential Mechanism of McSherry and Talwar~\cite{mcsherry2007mechanism}, a powerful tool from the differential privacy literature. We demonstrate that this mechanism is feasible and that it leads to a significant reduction in the adversary's success rate. In particular the exponential mechanism can  reduce the adversary's success rate by up to $12\%$ when $n$, the maximum number of hashing rounds, is two and by up to $18\%$ when $n=3$. 
%\vspace{-0.4cm}
\subsection{Constructing $\selpred$ via the Exponential Mechanism}
%\vspace{-0.2cm}
\label{subsec:The Exponential Mechanism}
To define the $\selpred$ function it suffices to specify the probability of each outcome $o \in \outputspace$ given an input password $pwd \in \pwdspace$. 
Consider the utility function $U(pwd,o) \doteq \frac{1-\stopcond(pwd,o)}{n-1} $ which assigns a utility score to each outcome $o \in \outputspace$ given a password $pwd$.  We can now use $U(pwd,o)$ to specify the probability of an outcome $o$ given password $pwd$. We define $\selpred_{exp,\epsilon}$ such that $$
\Pr[\selpred_{exp,\epsilon}(pwd) = o] \doteq \frac{e^{\epsilon U(pwd,o)}}{\sum_{o' \in \outputspace} e^{\epsilon U(pwd,o')}} \ .
$$
Intuitively, given a fixed password $pwd$ the exponential mechanism assigns a higher probability to outcomes $o$ with shorter stopping times  $\stopcond(pwd,o)$. 

We can ensure that our cost constraints are satisfied by tuning the parameter $k$ (the number of hash iterations per round). After $\epsilon$ and $\outputspace$ have been fixed we will set $k$ to be the maximum integer such that:
$$
\sum\limits_{i=1}^n iO_i \prob{i} = \frac{1}{W}\sum\limits_{i=1}^n iO_i e^{\epsilon U(i)} \le \alphak \ .
$$

where the left hand side is a constant.

Observe that whenever $\outputspace$ is symmetric it is possible to precompute the stopping time probabilities $\prob{j} \doteq \frac{\outputspace_j\cdot e^{\epsilon \cdot j}}{\sum_{i=1}^n \outputspace_i\cdot e^{\epsilon \cdot i} }$. Thus, to sample from $\selpred_{exp,\epsilon}(pwd)$ it suffices to sample the stopping time $j$, compute $\hash^{jk}(pwd,s_u)$ and sample uniformly at random from the set $\outputspace_{j,pwd}$. 

Theorem \ref{thm:diffPrivacy2} states that the exponential mechanism above satisfies our desired security constraint. Theorem \ref{thm:diffPrivacy2} is similar to the general result of McSherry and Talwar~\cite{mcsherry2007mechanism}, which would imply that our mechanism $\selpred_{exp,\epsilon}$ satisfies the security constraint with security parameter $2\epsilon$. In our particular setting we can exploit symmetry (e.g., $\left|\outputspace_{j,pwd'}\right|=\left|\outputspace_{j,pwd}\right|$) to obtain a tighter bound with security parameter $\epsilon$. %The proof of Theorem \ref{thm:diffPrivacy2} is in the appendix.

\newcommand{\thmDiffPrivacyTwo}{For any symmetric $\outputspace$ we have $$\forall pwd,pwd' \in \pwdspace, \forall o \in \outputspace, \frac{ \Pr[\selpred_{exp,\epsilon}(pwd) = o]}{\Pr[\selpred_{exp,\epsilon}(pwd') = o]} \le e^\epsilon \ .$$}
\begin{theorem} \label{thm:diffPrivacy2}
\thmDiffPrivacyTwo
\end{theorem}
\begin{proofof}{Theorem \ref{thm:diffPrivacy2}}
Let $w_{o,pwd} \doteq e^{\epsilon U(pwd,o)}$ denote the weight of outcome $o \in \outputspace$ given password $pwd$ and let $W_{pwd} \doteq \sum\limits_{o \in \outputspace} w_{o,pwd}$ denote the cumulative weight of all $o \in \outputspace$. Because $\outputspace$ is symmetric we have $W_{pwd} = W_{pwd'}$ for any pair of passwords $pwd,pwd' \in \pwdspace$. Thus, 
\confversion{
\begin{eqnarray*}
 \frac{\Pr\left[\mathbf{SelectPredicate}\left(pwd\right)=o\right]}{\Pr\left[\mathbf{SelectPredicate}\left(pwd'\right)=o\right]} 
&=& \frac{w_{o,pwd}/W_{pwd}}{w_{o,pwd'}/W_{pwd'}} \\  =  \frac{w_{o,pwd}}{w_{o,pwd'}} 
\leq e^{\epsilon\left|U(pwd',o)-U(pwd,o) \right|} &\leq& e^\epsilon \ .
\end{eqnarray*}}
\fullversion{
\[
 \frac{\Pr\left[\mathbf{SelectPredicate}\left(pwd\right)=o\right]}{\Pr\left[\mathbf{SelectPredicate}\left(pwd'\right)=o\right]} 
= \frac{w_{o,pwd}/W_{pwd}}{w_{o,pwd'}/W_{pwd'}}   =  \frac{w_{o,pwd}}{w_{o,pwd'}} 
\leq e^{\epsilon\left|U(pwd',0)-U(pwd,o) \right|} \leq e^\epsilon \ . \]
}

\end{proofof}

\cut{
Because we are working with a symmetric set $ \outputspace$ we can set define $U(j) \doteq U(pwd,o)$ whenever $o \in \outputspace_{j,pwd}$.  

 We note that since $\outputspace$ is symmetric} \cut{
\begin{equation*}
\begin{split}
W_{pwd} & = \sum\limits_{o \in \outputspace} e^{\epsilon U(pwd,o)}= \sum\limits_{i=1}^n O_i e^{\epsilon U'(i)}
\end{split}
\end{equation*}
and this quantity is independent of the password used. } \cut{ we have $W_{pwd} = W_{pwd'}$ for all passwords $pwd\ne pwd'$. Thus, we can set $W \doteq W_{pwd}$ for all $pwd \in \pwdspace$. }

 \cut{After $\epsilon$ and $\outputspace$ have been fixed we can adjust the parameter $k$ to ensure that this constraint holds. In particular, we will set $k$ to be the greatest integer such that the inequality holds. We therefore have that for a fixed security constant $\epsilon$, the exponential mechanism completely determines a feasible $\selpred$ algorithm.}

%\vspace{-0.7cm}
\subsection{Analysis of the Exponential Mechanism}
\begin{figure*}[t!]
    \centering
    \begin{subfigure}[t]{0.47\textwidth}
        \centering
        \includegraphics[width=1.1\textwidth]{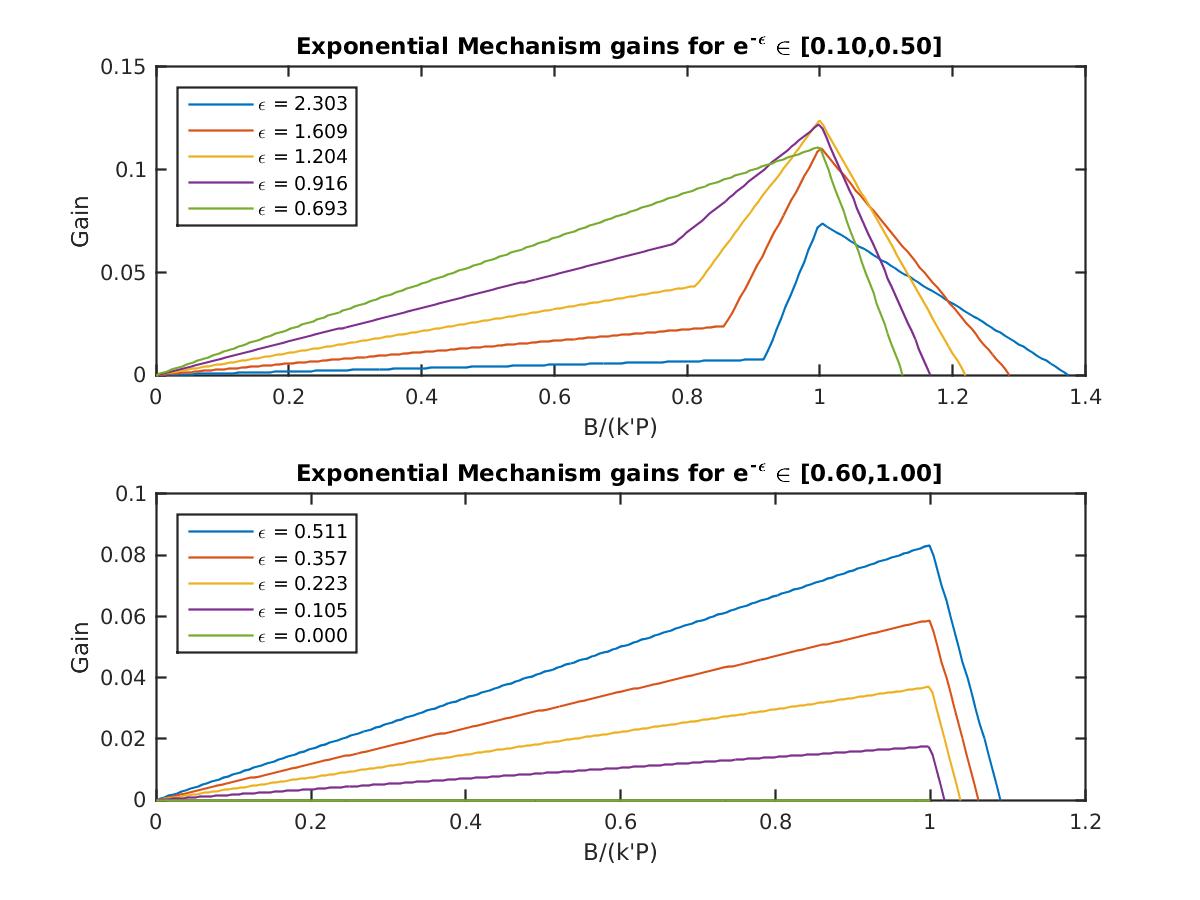}
\cvspace{-0.6cm}
        \caption{Two round exponential mechanism. }
                \label{subfig:twopred_exp}
    \end{subfigure}%
    ~ 
    \begin{subfigure}[t]{0.47\textwidth}
        \centering
        \includegraphics[width=1.1\textwidth]{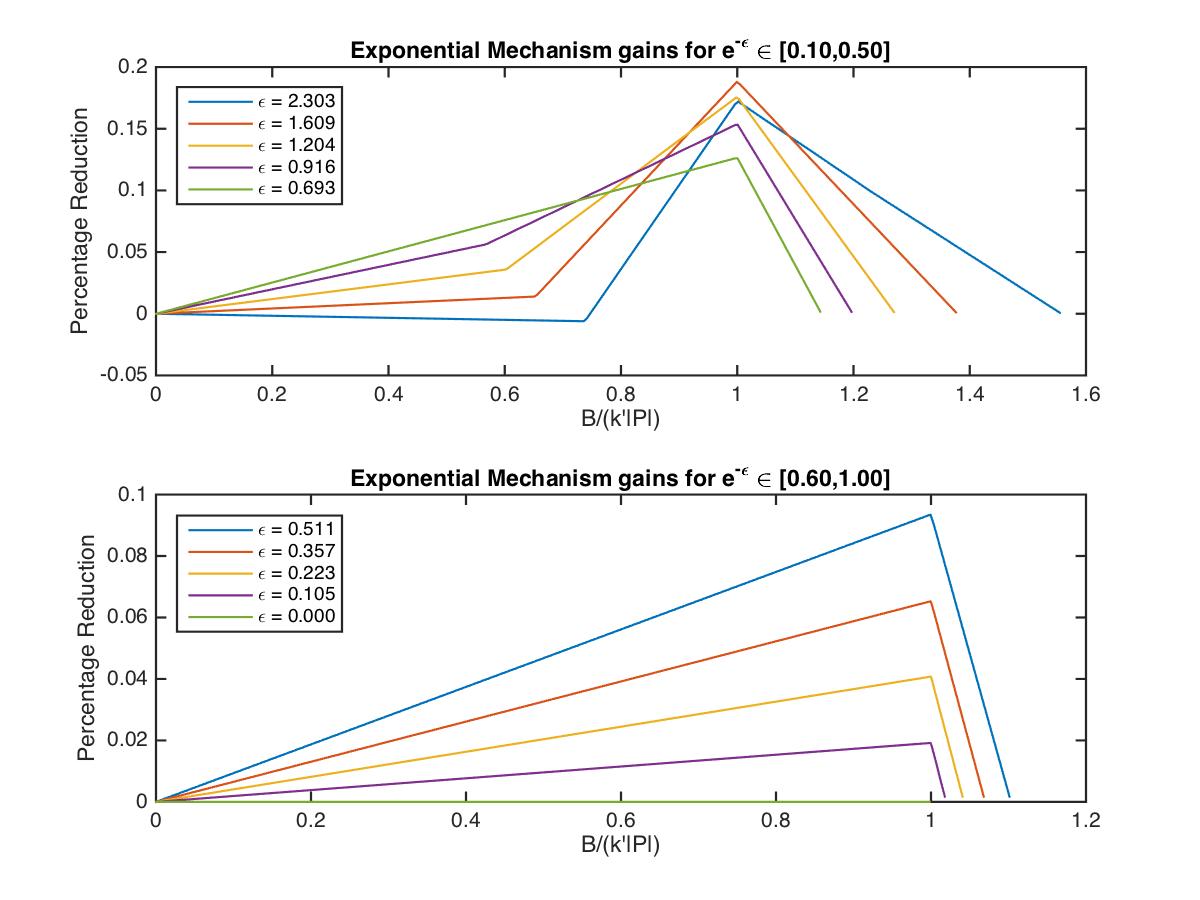}
        \cvspace{-0.6cm}
        \caption{Three round exponential mechanism. }
                \label{subfig:threepred_exp}
    \end{subfigure}
    ~
    \begin{subfigure}[t]{0.47\textwidth}
        \centering
        \includegraphics[width=1.1\textwidth]{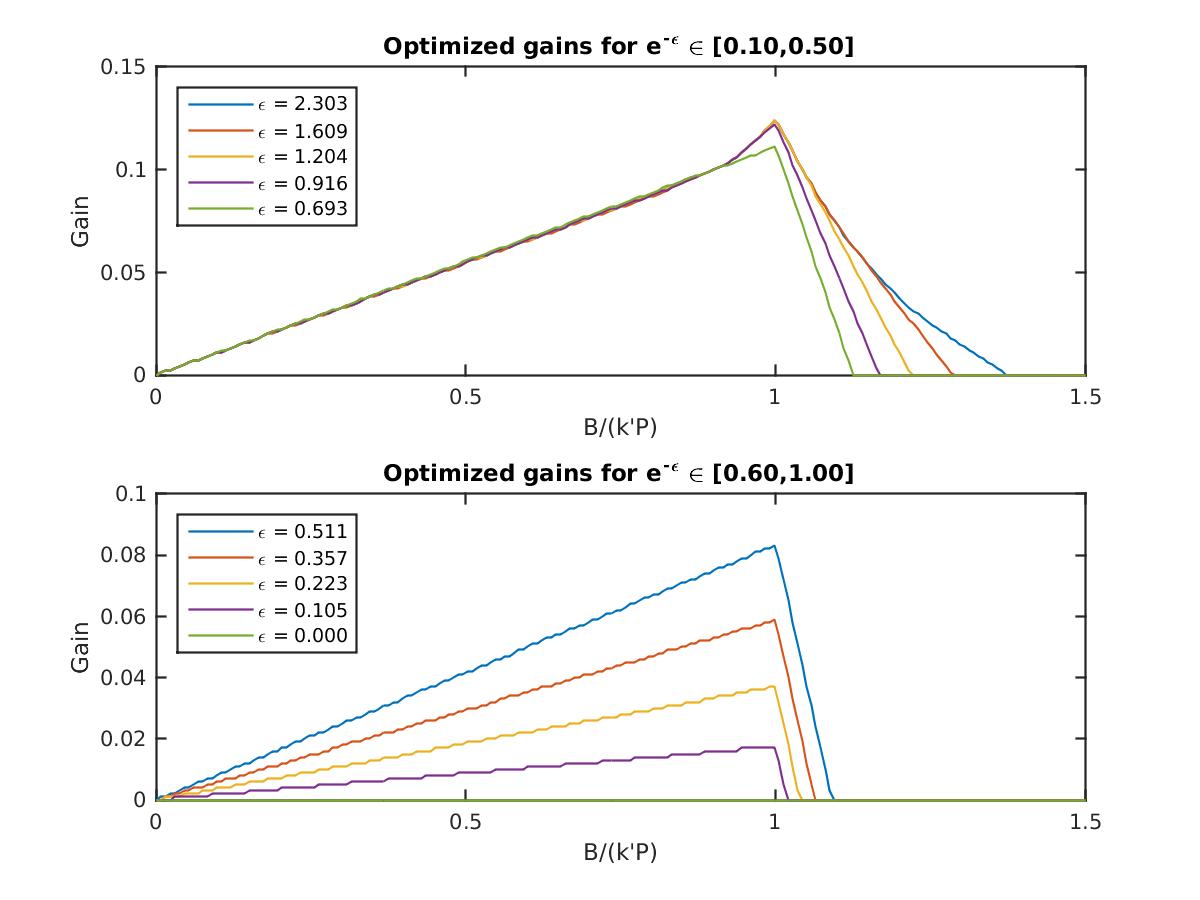}
\cvspace{-0.8cm}
        \caption{Two round optimal mechanism. }
                 \label{subfig:TwoRoundOpt}
    \end{subfigure}%
    ~ 
    \begin{subfigure}[t]{0.47\textwidth}
        \centering
        \includegraphics[width=1.1\textwidth]{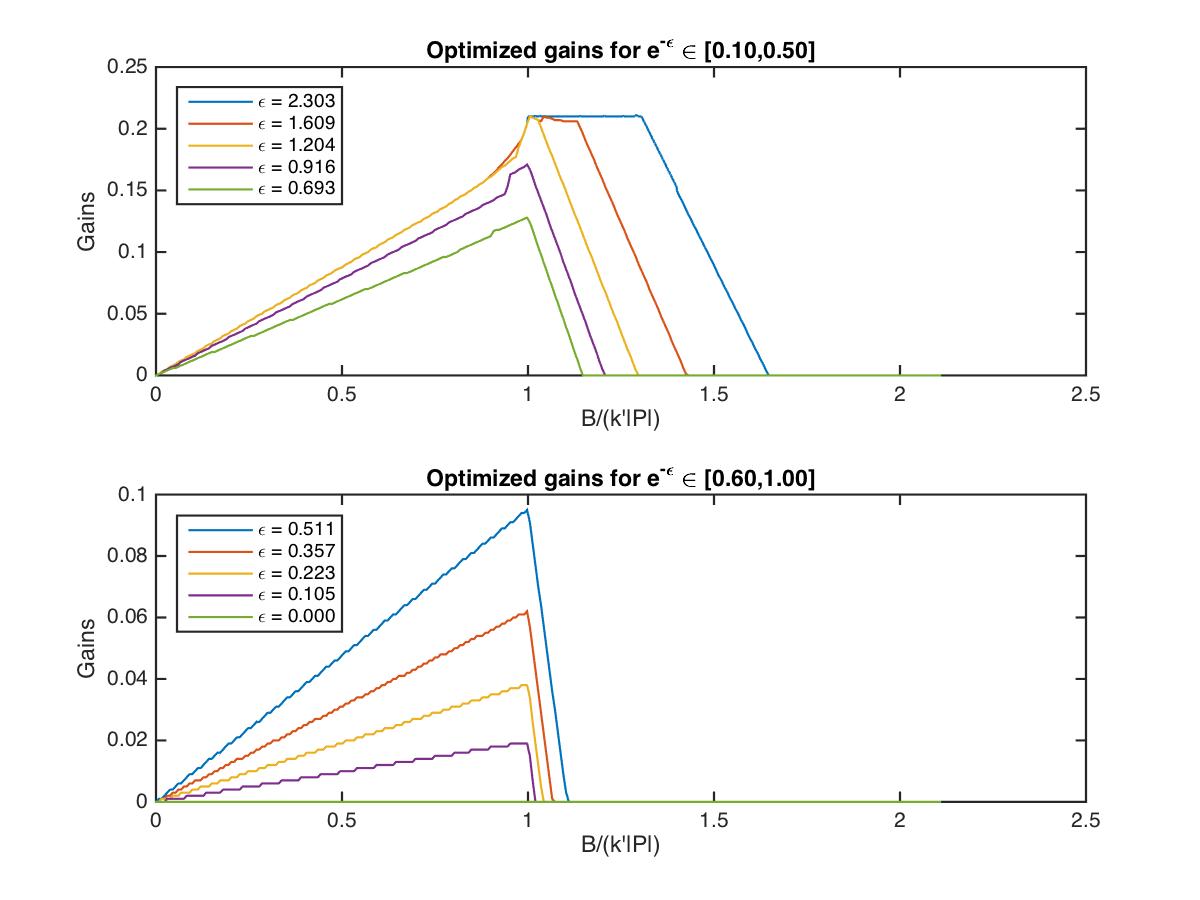}
        \cvspace{-0.8cm}
        \caption{Three round optimal mechanism. }
        \label{subfig:ThreeRoundOpt} 
    \end{subfigure}
    \cvspace{-0.2cm}
      \caption{Percent reduction in passwords breached $G_B = \pdetB-\padvB$ given as a function of the adversary's total budget $B$ with a normalizing factor of $\frac{1}{k'|\pwdspace|} = \frac{C_H}{C_{srv}|\pwdspace|}$ for various values of $\epsilon$. (a) and (b) show the two and three round exponential mechanisms, respectively, while (c) and (d) show the optimal gains for the two and three round mechanisms when the adversary's budget is known in advance.   }
     \label{fig:exponential}
     \cvspace{-0.3cm}
\end{figure*}

%\vspace{-0.2cm}
% !TEX root = ASIA_CCS_PWD_SYSTEM_2016.tex
% 
% 
We will use $G_B \doteq \pdetB-\padvB$ to denote the gains from adopting \clientcash~(i.e., the reduction in the probability that the adversary cracks the password). Recall that $P_{adv,B}$ (resp. $P_{det,B}$) denotes the probability that the optimal offline adversary successfully cracks a password protected with \clientcash~(resp. deterministic key-stretching) with a budget $B$.  Figure  \ref{subfig:twopred_exp} and \ref{subfig:threepred_exp} plots $G_B$ as a function of $B$ for various values of the security parameter $\epsilon$.  Theorem \ref{thm:fullpadv} allows us to compute $\padvB$ and $G_B$ efficiently. 

 \subsubsection{Two-Round Exponential Mechanism}
 Note that when $n=2$, $\outputspace = \{P_{0,2},P_{1,2}\}$ so $(O_1,O_2) = (1,1)$. Thus for a fixed value of $\epsilon$, we can define the exponential mechanism as follows: 
\begin{equation*}
\begin{split}
W  = 1 + e^{-\epsilon}\ , & ~~~~\prob{1} = \frac{1}{W} \ , ~~~~ \prob{2}  = \frac{e^{-\epsilon}}{W} \\
\end{split}
\end{equation*}
\cvspace{-0.25cm}
\begin{equation*}
\begin{split}
\hprob{1} + \hprob{2} &= \prob{1} + \prob{2} = 1 \\
\hprob{1} + 2\hprob{2} &= \prob{1} + 2\prob{2}  \le \alphak \\
\end{split}
\end{equation*}
Although we have an inequality in the last equation, we emphasize that $k$ is determined to be the maximum integer such that the inequality holds, as described in \textsection $\ref{subsec:The Exponential Mechanism}$. Since $k$ is large, and thus marginal changes in $\alphak$ are small, we assume for simplicity that $\prob{1} + 2\prob{2} = \alphak$ and round $k$ downwards if it is not an integer. 
 \subsubsection{Three-Round Exponential Mechanism}
 
 When $n=3$, $\outputspace = \{P_{0,3},P_{1,3},P_{2,3}\} \times \{P_{0,3},P_{1,3},P_{2,3}\}$ so $(O_1,O_2,O_3) = (3,2,4)$. Thus for a fixed $\epsilon$, we can define the exponential mechanism as follows: 
% Note: \prob{1} = 1/W instead of 3/W because \prob{1} denotes the probability of a single outcome with stopping time 1, \hprob{1}=3/W denotes the outcome of all outcomes with stopping time 1.
\begin{equation*}
\begin{split}
W & = 3 + 2e^{-\epsilon/2} + 4e^{-\epsilon}\\
(\prob{1},\prob{2},\prob{3}) & = \left( \frac{1}{W},\frac{e^{-\epsilon/2}}{W}, \frac{e^{-\epsilon}}{W}  \right)\\
\hprob{1} + \hprob{2} + \hprob{3} & = 3\prob{1} + 2\prob{2} + 4\prob{3} = 1 \\
\hprob{1} + 2\hprob{2} + 3\hprob{3} & = 3\prob{1} + 4\prob{2} + 12\prob{3} \leq \alphak
\end{split}
\end{equation*}

\paragraph{Discussion} Our analysis shows that exponential mechanism can be used to reduce the percentage of passwords breached by an optimal offline adversary. When $B = k'|\pwdspace|$ we will always have $\pdetB = 1$. Thus, we expect that the gain $G_B$ will decrease monotonically after this point which is indeed what we observe in the plots. In the two round case we always had positive gain (e.g., for all $B\geq 0$ we had $G_B > 0$ for every value of  $\epsilon$ that we tried). In the three round case we always had $G_B >0$ whenever $\epsilon < 2.3$. Observe that $G_B$ does not always increase monotonically as we increase $\epsilon$ (e.g., as we relax the security constraint). In both the two round and the three round case $G_B$ increases monotonically with $\epsilon$ as long as $\epsilon \le 0.5$, but after this the graphs intersect  indicating that different values of $\epsilon$ are better than others for different $B$ when the exponential mechanism is used. 

We note that the exponential mechanism does not incorporate any information about the adversary's budget $B$. The exponential mechanism $\selpred_{exp,\epsilon\approx2.3}$ with the largest value of $\epsilon$ performs poorly against smaller budgets $B$, but it provided the optimal defense in two rounds against an adversary with larger budgets. If we know $B$ in advance then we can often improve our construction of the $\selpred$ function (see \textsection \ref{sec:Client CASH Security Optimization}). 

\cut{
  Another observation is that for $\epsilon \le 0.5$ in both the two round and three round case, the reduction in password percentage is monotonically increasing in $\epsilon$ for all values of $B$. However, for greater $\epsilon$ several of the graphs intersect, indicating that different values of $\epsilon$ are better than others for different $B$ when the exponential mechanism is used. This suggests that at least in some cases, the exponential mechanism does not give optimal security - otherwise, increasing the security parameter $\epsilon$ will give a reduction in passwords that is at least as good as a smaller $\epsilon$. 

\paragraph{Discussion. } The exponential mechanism already reduces the adversary's success rate.  }

%\vspace{-0.6cm}
\section{\clientcash~ Optimization}
%\vspace{-0.35cm}
\label{sec:Client CASH Security Optimization}
% !TEX root = ASIA_CCS_PWD_SYSTEM_2016.tex
In this section we show how to optimize the $\selpred$ function under the assumption that the size of the password space $\left|\pwdspace\right|$ and the adversary's budget $B$ are known in advance. This may not be an unreasonable assumption for many users. For example, Bonneau and Schechter estimated that for the SHA256 hash function $C_H = \$7 \times 10^{-15}$~\cite{BS14}, and Symantec reports that passwords generally sell for $\$ 4 - \$30$ on the black market. Thus, we might reasonably expect that a rational adversary will have a budget $B \leq \frac{30}{7 \times 10^{-15}} = 4.29 \times 10^{15}$. If the user selects passwords uniformly at random then it is reasonable to assume that $\left|\pwdspace\right|$ is known. For example, we might have $|\pwdspace| = 10^{12}$ for users who memorize three random person-action-object stories and use the associated action-object pairs for their password~\cite{blockiNaturallyRehearsingPasswords,spacedRepetitionAndMnemonics}. 

\cut{
 Assuming that $k' = 10^7$ and that a user uses a password scheme similar to Naturally Rehearsing Password Scheme \anote{cite} to obtain $|\pwdspace| = 10^{12}$, we can simply use the optimal distribution data for $\frac{B}{k'|\pwdspace|} = \frac{4.29 \times 10^{15}}{10^19} = 4.29 \times 10^{-7}$. However, if the adversary's budget is not known, the user may implement the exponential distribution in $\selpred$ since a fixed value of $\epsilon$ determines all parameters. For the choices of $\epsilon$ examined, the exponential mechanism reductions coincides with or comes very close to the optimal reductions, which makes it a useful tool for designing $\selpred$ when we are uncertain about $B$.  }

Given a symmetric set $\outputspace$ of outcomes our goal is to find probability values $\prob{1},\ldots,\prob{n}$ and a parameter $k$ which minimize the adversary's success rate $\padvB$ in the event of an offline attack. We must choose these values in a way that satisfy our security and cost constraints (Table \ref{tab:Constraints}). This goal is stated formally as Optimization Goal \ref{goal:MinPadv}.

\setlength{\FrameSep}{0.01cm} 
\begin{goal}
\begin{framed}
%\vspace{-0.2cm}

\noindent {\bf Input Parameters: } $C_{srv}, C_H, \epsilon, \outputspace, B, \left|\pwdspace\right|, n$ \newline
\noindent {\bf Variables: } $\prob{1}, \ldots, \prob{n}, k, \padvB$ \newline
\noindent {\bf minimize } $\padvB$ subject to
 \cvspace{-0.4cm}
\confversion{\begin{align*} 
(1) & \forall i \in \{1,\ldots, n\},~~\prob{i} \in [0,1] &  (2) & \sum\limits_{i=1}^n O_i\cdot\prob{i} = 1 \\ 
(3) & \forall i,j \in \{1,\ldots,n\}, ~~ \frac{\prob{i}}{\prob{j}} \le e^{\epsilon} ~~~~~~~& (4)& k \in \left\{1,\ldots,\frac{C_{srv}}{C_H} \right\} \\
\end{align*}
\cvspace{-1.1cm}
\begin{align*} 
(5) & \padvB \geq \pwdratio \max\limits_{\mathbf{b} \in F^*}\left\{
O_1\prob{1}b_1 + \sum\limits_{i=2}^n O_i\prob{i}b_i \prod\limits_{j=1}^{i-1} \left(\frac{\ell_j}{\ell_j - 1} \right)
\right\} & & \\
 (6)  &  \sum\limits_{j=1}^n jO_j\prob{j} \le \frac{C_{srv}}{kC_H} & & 
 \end{align*}}
 \fullversion{
 \begin{align*} 
 (1) ~& \forall i \in \{1,\ldots, n\},~~\prob{i} \in [0,1] &  (2)~ & \sum\limits_{i=1}^n O_i\cdot\prob{i} = 1 \\ 
(3) ~& \forall i,j \in \{1,\ldots,n\}, ~~ \frac{\prob{i}}{\prob{j}} \le e^{\epsilon} ~~~~~~~& (4)~ & k \in \left\{1,\ldots,\frac{C_{srv}}{C_H} \right\} \\
(5) ~& \padvB \geq \pwdratio \max\limits_{\mathbf{b} \in F^*}\left\{
O_1\prob{1}b_1 + \sum\limits_{i=2}^n O_i\prob{i}b_i \prod\limits_{j=1}^{i-1} \left(\frac{\ell_j}{\ell_j - 1} \right)
\right\} & (6)~ &  \sum\limits_{j=1}^n jO_j\prob{j} \le \frac{C_{srv}}{kC_H} 
 \end{align*}
}
\cvspace{-0.2cm}
\end{framed}
\cvspace{-0.4cm}
\caption{Minimizing the adversary's probability of success. }
\label{goal:MinPadv}

\end{goal}

Constraints (1) and (2) specify that the $\prob{i}$ values must define a valid probability distribution and constraints (3) and (6) are the security and cost constraints from Table \ref{tab:Constraints}. Constraint (4) states that $k$, the parameter controling the cost of the underlying hash function $\hash_k$ in each round of \clientcash, can be at most $\frac{C_{srv}}{C_H}$. Otherwise it would be impossible to satisfy our cost constraint because just a single round of hash iteration would cost more than $C_{srv}$. Finally, constraint (5) implies that we must minimize $\padvB$ subject to the assumption that the adversary responds with his optimal strategy (see Theorem \ref{thm:fullpadv} ). 

We note that, for a fixed value of $k$, Optimization Goal \ref{goal:MinPadv} becomes a linear program. Thus, we can solve Optimization Goal \ref{goal:MinPadv} efficiently by solving the resulting linear program for each value of $k \in \{1,\ldots,\frac{C_{srv}}{C_H}\}$ and taking the best solution.

\begin{figure*}[t!]
    \centering
    \begin{subfigure}[t]{0.5\textwidth}
        \centering
        \includegraphics[width=1.1\textwidth]{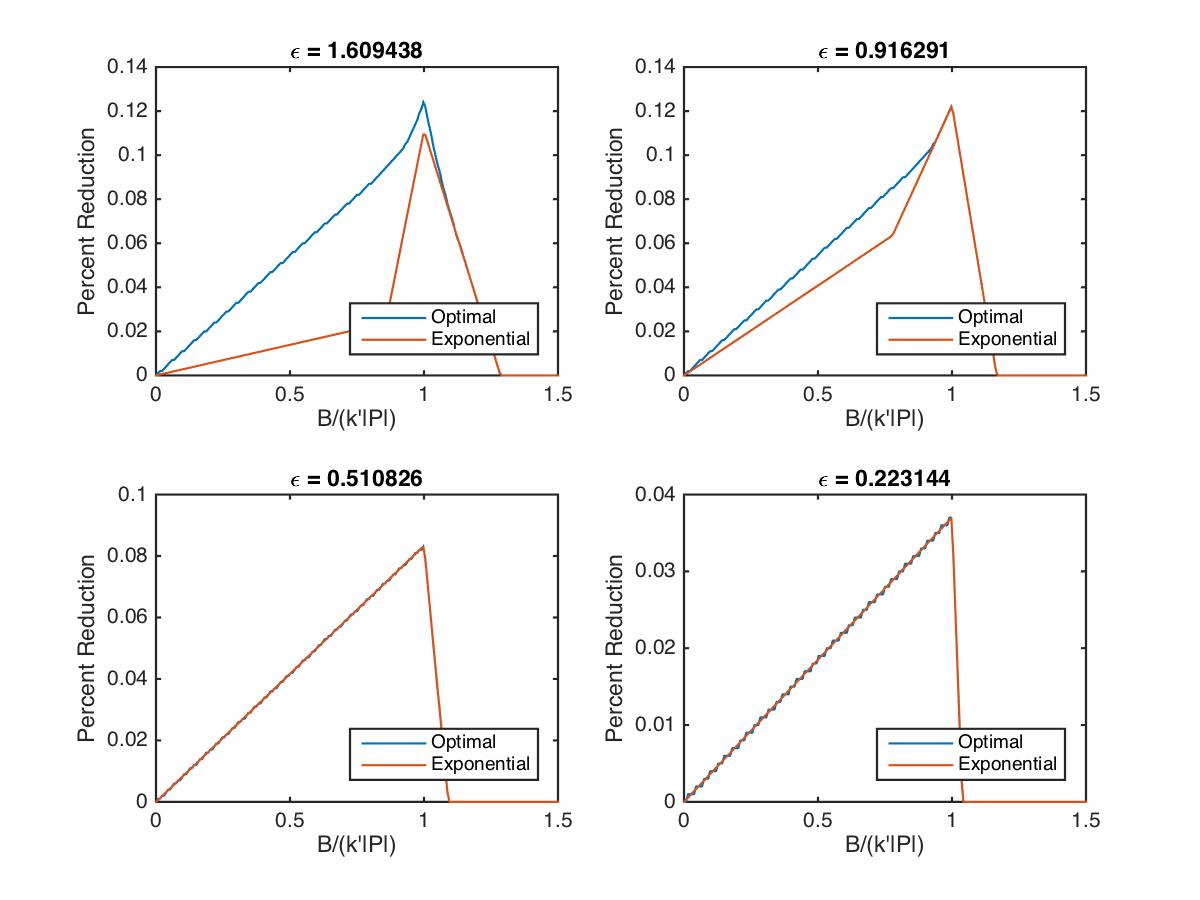}
\cvspace{-.50cm}
        \caption{Two round mechansims. }
\label{subfig:TwoRoundOptVsExp}
    \end{subfigure}%
    ~ 
    \begin{subfigure}[t]{0.5\textwidth}
        \centering
        \includegraphics[width=1.1\textwidth]{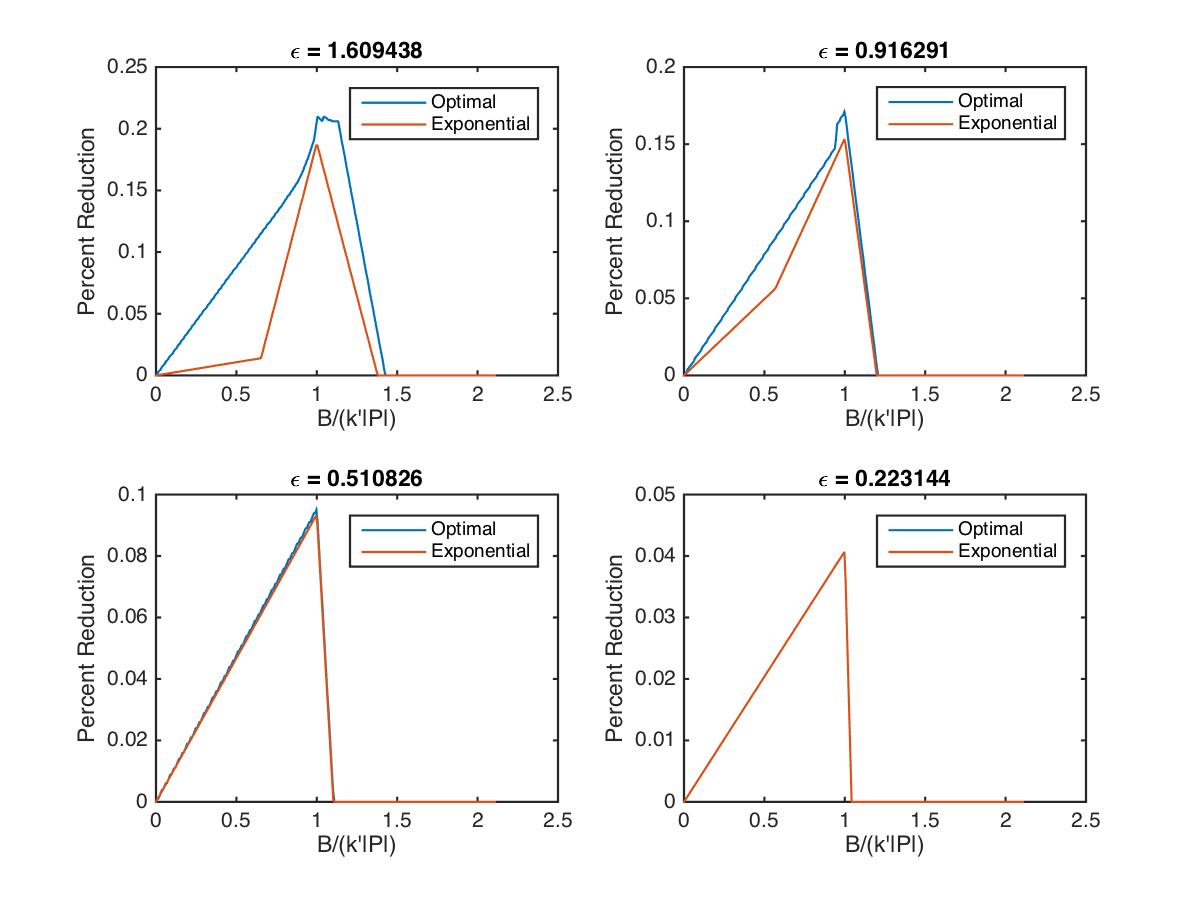}
\cvspace{-.50cm}
        \caption{Three round mechanisms. }
\label{subfig:ThreeRoundOptVsExp}
    \end{subfigure}
    \caption{Comparison of the percent reduction in cracked passwords $G_B$ between the optimal distribution (blue) and the exponential distribution (orange) for as a function of $B$ with a normalizing factor of $\frac{1}{k'|\pwdspace|} = \frac{C_H}{C_{srv}|\pwdspace|}$ for various values of $\epsilon$. The results for the two round optimal mechanism is shown in (a), while the three round optimal mechanism is shown in (b).}
\cvspace{-0.4cm}
\end{figure*}

 \cut{Notice that in a traditional password hashing system, the hash should be iterated as many times as possible to increase costs for the adversary, so $k'$ is the largest integer such that $k' \times C_H \le C_{srv} \Rightarrow k' \le \frac{C_{srv}}{C_H}$. $k$ cannot be greater than $k'$, since we do not change $C_{srv}$ and we hash iterations that are integer multiples of $k$. Thus, if we assume for simplicity that $k' = \frac{C_{srv}}{C_H}$, constraint (2) follows. Constraint (3) follows from the fact that $\padv$ is a probability, and constraint (7) follows from Corollary \ref{cor:padv} in \textsection \ref{sec:The Adversary}. }

\subsection{Analysis} 
 Figures \ref{subfig:TwoRoundOpt} and \ref{subfig:ThreeRoundOpt} compare the optimal \clientcash~defense to the cost-equivalent deterministic key-stretching defense. We use the same symmetric predicate sets  $\outputspace$ as in previous sections (two-round and three-round). \confversion{The full version of this paper~\cite{fullVersion} includes additional figures comparing the exponential mechanism and the optimal mechanism directly.}\fullversion{Figures \ref{subfig:TwoRoundOptVsExp} and \ref{subfig:ThreeRoundOptVsExp} in the appendix directly compare the exponential mechanism and the optimal mechanism.} 

\cvspace{-0.18cm}
\paragraph{Discussion} Our results demonstrate that it is often possible to obtain even greater reductions in the adversary's success rate if we know the budget $B$ in advance. For example, in the three round case $n=3$ we can reduce the adversary's rate of success by as much as $G_B = 21\%$.  However, we note that the solution to Optimization Goal \ref{goal:MinPadv} may not be optimal if the estimate of the adversary's budget $B$ is wrong. Thus, if the user is uncertain about the adversary's budget he may better off adopting the exponential mechanism. 

Unlike the exponential mechanism we expect for the gain $G_B$ to increase monotonically with larger $\epsilon$ because we only loosen constraint (3). This is what we observe in Figures \ref{subfig:TwoRoundOpt} and \ref{subfig:ThreeRoundOpt}. We also note that when the adversary's budget $B$ exceeds $k'|\pwdspace|$ we have $\pdetB = 1$. Thus, $G_B$ must decrease monotonically after this point because $\padvB$ will increase monotonically with $B$. 

\cut{
This is clearly the case (see Figures \ref{subfig:TwoRoundOpt} and \ref{subfig:ThreeRoundOpt}). For $\epsilon \le 0.5$, the reduction is strictly increasing as a function of $\epsilon$; however as $\epsilon$ increases further in both cases, the reduction in passwords breached as a function of $B$ seems to converge to some limiting distribution which has a maximum reduction in password breaches of $12\%$ when $n=2$ and $21\%$ when $n=3$. It is also important to note that, if the adversary's budget $B$ exceeds $k'|\pwdspace|$, the adversary's probability of success with deterministic hashing is $P_{det} = 1$, since the adversary can apply $\hash^{k'}$ to every $pwd \in \pwdspace$. Thus, after the point where $\frac{B}{k'|\pwdspace|} = 1$, the optimal graphs must be non-increasing.    \jnote{Explain that after $B/(k'|P|)$ reaches $1$ the adversary achieves $100\%$ success rate against the deterministic mechanism. Thus, the gain can only drop after this point. Add explanation of ``suspicious looking" jumpy behavior in 3 predicate case. }}
\cut{
\begin{figure*}[t!]
    \centering
    \begin{subfigure}[t]{0.5\textwidth}
        \centering
        \includegraphics[width=1.1\textwidth]{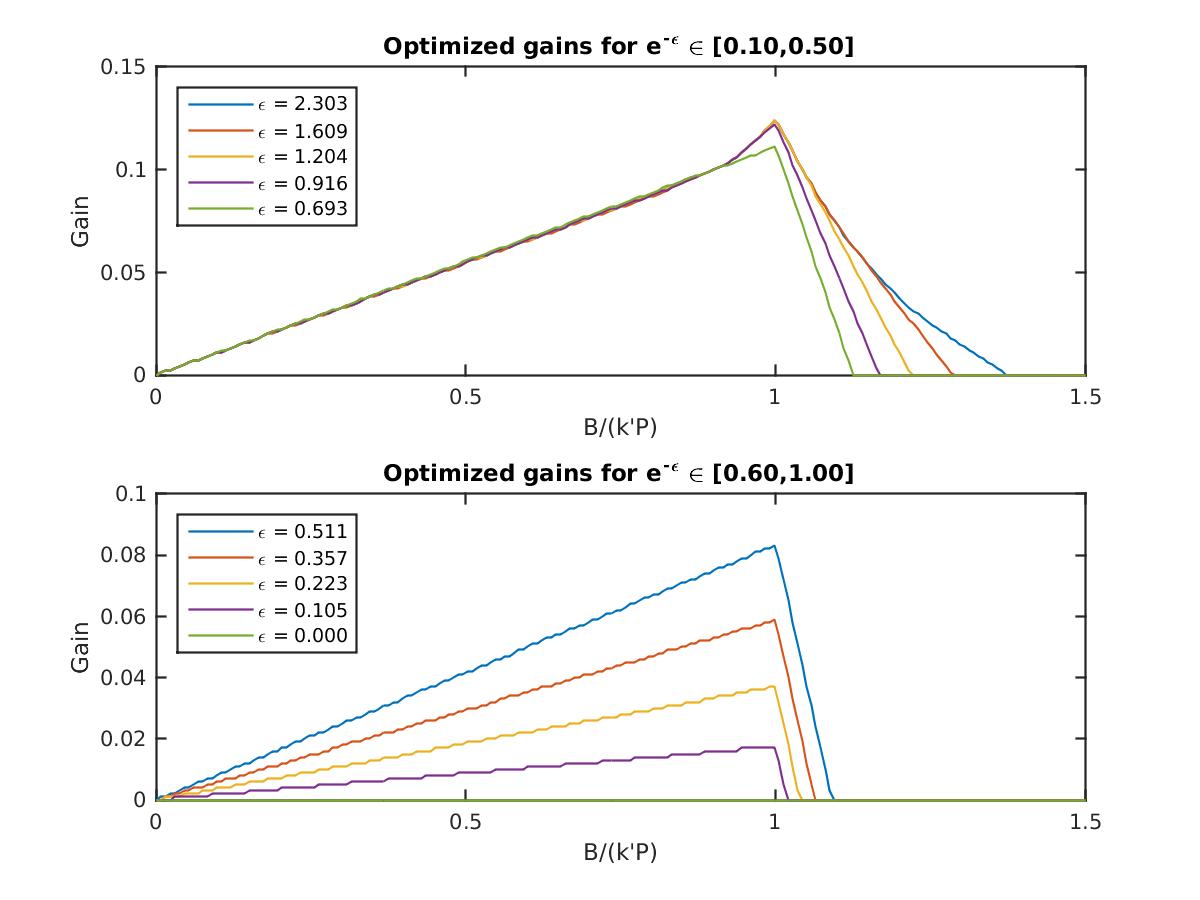}
        \caption{Two round optimal mechanism. }
 \label{subfig:TwoRoundOpt}
    \end{subfigure}%
    ~ 
    \begin{subfigure}[t]{0.5\textwidth}
        \centering
        \includegraphics[width=1.1\textwidth]{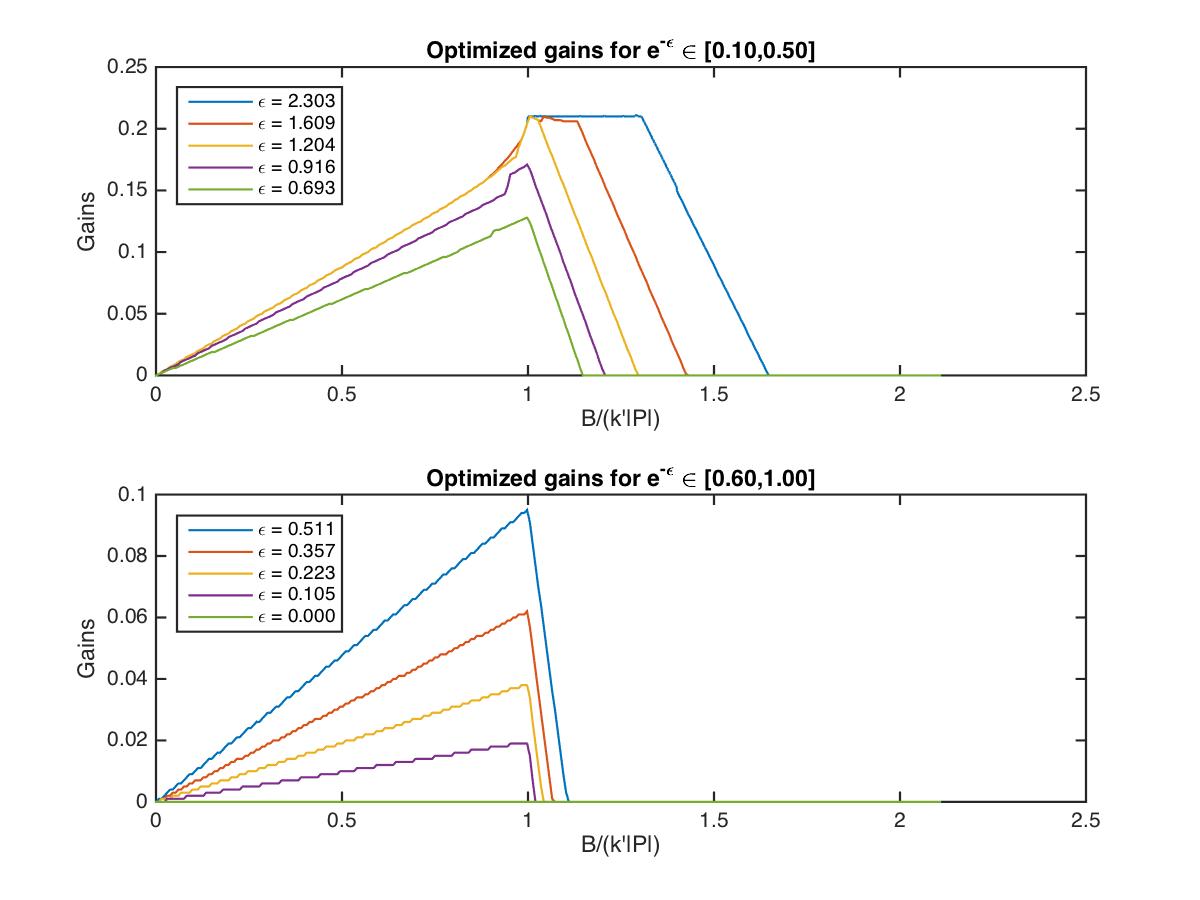}
        \caption{Three round optimal mechanism. }
\label{subfig:ThreeRoundOpt}
    \end{subfigure}
    \caption{Percent reduction in passwords breached given as a function of the adversary's total budget $B$ with a normalizing factor of $\frac{1}{k'|\pwdspace|} = \frac{C_H}{C_{srv}|\pwdspace|}$ for various values of $\epsilon$. The results for the two round optimal mechanism is shown in (a), while the three round optimal mechanism is shown in (b).  }
\end{figure*}
}
\cut{
\paragraph{Comparison with the Exponential Mechanism. } As shown in Figures \ref{fig:twopred_cmp} and \ref{fig:threepred_cmp}, the optimal percent reductions (blue) are at least as good as the Exponential Mechanism reductions for each value of $\epsilon$ shown, which is expected of the optimal distributions.When $\epsilon \le 0.5$, the optimal reductions match that of the exponential mechanism reductions. However, as $\epsilon$ increases past this amount, the differences are much more pronounced. Even so, for $n=2$ the exponential mechanism distribution coincides with the optimal distribution for some ranges of $B$, and the same is true for $n=3$ when $\epsilon \le 0.92$. However, in Figure \ref{fig:threepred_cmp}, for $\epsilon = 1.609$, the optimal percent reductions are strictly greater than the reductions given by the exponential mechanism. We also note that the exponential mechanism is completely determined for a fixed $\epsilon$, whereas we treat $k$ as a variable in Optimization Goal \ref{goal:MinPadv}. If we know the adversary's budget within a certain degree of accuracy, we can use the data from Figures \ref{fig:twopred_opt} and \ref{fig:threepred_opt} to guide our choice of distribution and $k$.}
\cut{
\paragraph{Two-Round vs Three-Round \clientcash}  In general, the three-round mechanism outperforms the two-round mechanism. This suggests that we may be able to obtain even greater reductions $G_B$ in the adversary's success rate by moving to a four round mechanism $n=4$. However, we note that larger values of $n$ will not necessarily be superior due to our differential privacy constraint. For example, the differential privacy constraint ensures outcomes with stopping time $\stopcond(pwd,o)=n$ will be output with comparable probability to outcomes $o'$ with short stopping times.}

%\vspace{-0.5cm}
%
\section{Practical Guidelines} \label{sec:practicalguideline}
% !TEX root = ASIA_CCS_PWD_SYSTEM_2016.tex

All the \clientcash~ mechanisms examined give reductions in the percentage of passwords breached by an optimal offline adversary. We also make the following recommendations of when a user should use a certain mechanism. 

\begin{figure*}[t!]
    \centering
    \begin{subfigure}[t]{0.5\textwidth}
        \centering
        \includegraphics[width=1.1\textwidth]{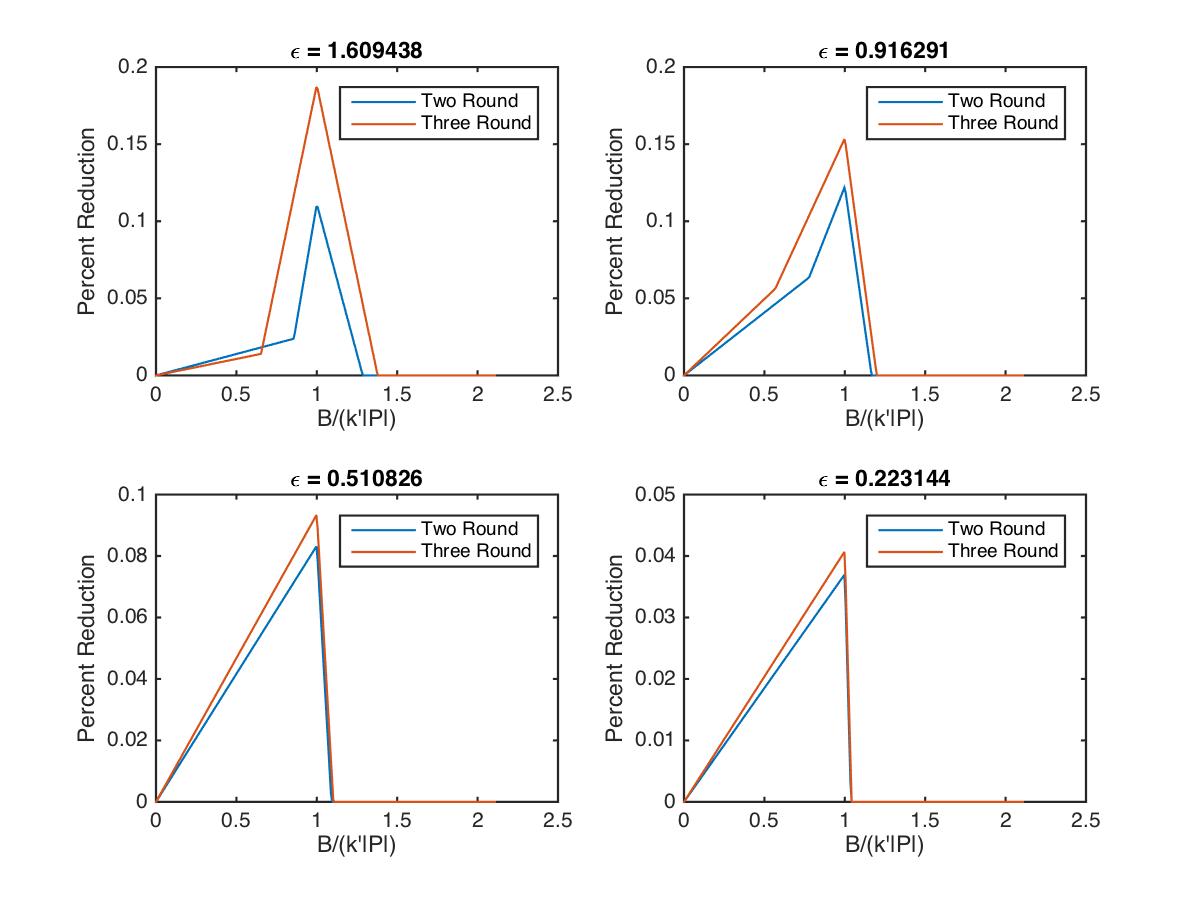}
        \caption{Comparison of exponential mechanisms. }
        \label{subfig:expcmp}
    \end{subfigure}%
    ~ 
    \begin{subfigure}[t]{0.5\textwidth}
        \centering
        \includegraphics[width=1.1\textwidth]{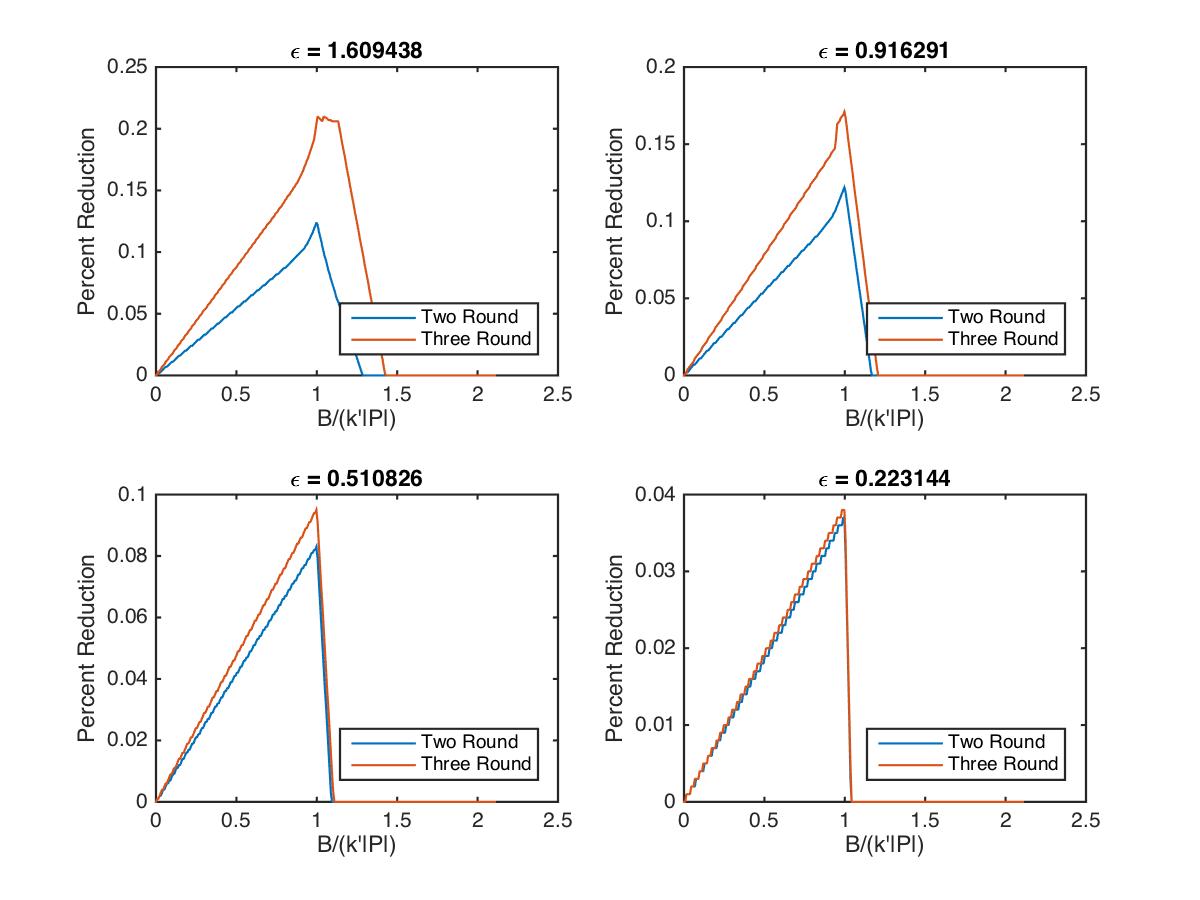}
        \caption{Comparison of optimal mechanisms. }
        \label{subfig:optcmp}
    \end{subfigure}
    \caption{
    Comparison of the percent reduction in passwords between two-round mechanisms (blue) and three-round mechanisms (orange) as a function of $B$ with a normalizing factor of $\frac{1}{k'|\pwdspace|} = \frac{C_H}{C_{srv}|\pwdspace|}$ for various values of $\epsilon$. The results for the exponential mechanism are shown in (a), while the optimal mechanism is shown in (b).  }
\label{fig:cmpfig}
\end{figure*}
\cvspace{-0.18cm}
\paragraph{Two Rounds vs. Three Rounds} Figures \ref{subfig:expcmp} and \ref{subfig:optcmp} compare the two-round \clientcash~mechanism mechanism to the three-round \clientcash~mechanism. Based on the results in Figure \ref{fig:cmpfig}, we recommend using three rounds of hashing in general. In particular, the three round mechanisms tend to achieve greater gains in security. The one possible exception is when the adversary's budget is smaller, but we are uncertain about the adversary's exact budget. In this one case the two round exponential mechanism might be the right choice~\footnote{In Figure \ref{subfig:expcmp}, the two round exponential mechanism will outperform the three round mechanism when $\frac{B}{k'|\pwdspace|} \le 0.5$  for security parameters $\epsilon \ge 1.6$.}. Against adversaries with a larger budget the three round mechanism is the clear winner --- note that the gains remain positive for much larger values of the adversary budget $B$. 
\confversion{\cvspace{-0.18cm}}
\paragraph{Information Leakage} In general we would recommend that any implementation use $\epsilon \le 2.08$ so that the outcome $o$ leaks at most $3$ bits of information about the user's password. We note that if the adversary breaches only the client then he would still need to complete a brute-force search over all passwords  $pwd \in \pwdspace$ to extract this information from $o$. In our demo implementation we selected security parameter $\epsilon =  1.609$ so that the outcome $o$ of $\selpred$ will leak at most $2.32$ bits of information about the user's password.  Future versions might help the client to tune $\epsilon$ himself. We selected $\epsilon =  1.609$ because it yielded the highest reductions against an adversary with a larger budget $B$. 
\cvspace{-0.18cm}
\paragraph{Adversary Budget} If we have an accurate estimation of the adversary's budget $B$ then we should compute the $\selpred$ distribution using our algorithm from \textsection \ref{sec:Client CASH Security Optimization} before implementing \clientcash. However, if this estimation is not precise then it may be better to use the exponential mechanism from \textsection 6.
\cvspace{-0.18cm}
\paragraph{Password Distribution} We only show how to optimize \clientcash~under the assumption that users choose passwords uniformly at random. We hope that future work will extend this analysis to more general password distributions. Until then \clientcash~should only be adopted by users who select uniformly random passwords from some space. 
\cvspace{-0.18cm}
\paragraph{Amortization of Costs} Our guarantee that the client's computational costs during authentication at most $C_{srv}$ only holds in expectation. Of course the user may get unlucky when we run $\selpred(pwd)$ and end up with the maximum stopping time $\stopcond(pwd,o)=n$. We note that if key-stretching was performed by a trusted third party (e.g., LastPass) then these costs could be amortized over different users. If key-stretching is only performed on the user's local computer then we could give an unlucky user the option to speed up the key-stretching process by memorizing an additional random character and appending it to his password (e.g., see \cite{BS14}).

\subsection{Demo Implementation}  \label{sec:Implementation}
%\vspace{-0.4cm}
% !TEX root = ASIA_CCS_PWD_SYSTEM_2016.tex
A demo implementation of the three round exponential \clientcash~mechanism is available on {\color{blue}\underline{client-cash.github.io}}\footnote{The current demo implementation does not permanently save state for future authentication sessions though we anticipate adding this functionality soon.}. We implemented our algorithms in Javascript so the key-stretching is done locally by the user's web browser. In our implementation we used $k=100,000$ iterations of the SHA256 hash function $\hash$. Thus, \clientcash~perform between $100,000$ and $300,000$ iterations of the SHA256 hash function to derive our final password. We selected $n=3$ because the exponential mechanism performs better with $n=3$ rounds of hashing (see Figure \ref{subfig:expcmp}).

\cut{
Any implementation of \clientcash~should consider the following points: 

\noindent{\em Information Leakage. } In our implementation we selected security parameter $\epsilon =  1.609$. Future versions might help the user to tune these parameters himself. We selected $\epsilon =  1.609$ because it yielded the highest reductions against an adversary with a larger budget $B$. The outcome $o$ of $\selpred$ can leak at most $2.32$ bits of information about the user's password by itself. If the adversary breaches only the client  would need to complete a brute-force search over all passwords  $pwd \in \pwdspace$ to extract this information from $o$. In general we would recommend that any implementation use $\epsilon \le 2.08$ so that the outcome $o$ leaks at most $3$ bits of information about the user's password. 

\noindent{\em Password Distribution.} We only show how to optimize \clientcash~under the assumption that users choose passwords uniformly at random. We hope that future work will extend this analysis to more general password distributions. Until then \clientcash~should only be adopted by users who select uniformly random passwords from some space. 

\noindent{\em Adversary Budget.} If we have an accurate estimation of the adversary's budget $B$ then we should compute the $\selpred$ distribution using our algorithm from \textsection \ref{sec:Client CASH Security Optimization} before implementing \clientcash. However, if this estimation is not precise then it may be better to use the exponential mechanism from \textsection \ref{sec:Analysis of the Exponential Mechanism}.

\noindent{\em Amortization of Costs.} Our guarantee that the client's computational costs during authentication at most $C_{srv}$ only holds in expectation. Of course the user may get unlucky when we run $\selpred(pwd)$ and end up with the maximum stopping time $\stopcond(pwd,o)=n$. We note that if key-stretching was performed by a trusted third party (e.g., LastPass) then these costs could be amortized over different users. If key-stretching is only performed on the user's local computer then we could give an unlucky user the option to speed up the key-stretching process by memorizing an additional random character and appending it to his password (e.g., see \cite{BS14}).
}
\cut{

\begin{itemize}
\item {\bf Password distribution. }The mechanisms analyzed have been optimized with the assumption that users choose secure passwords; that is, passwords are selected uniformly at random from a fixed space of passwords $\pwdspace$. \clientcash~may be used in this way to extend the functionalities of password management schemes like LastPass or Naturally Rehearsing Passwords, which uniformly selects passwords. \clientcash~may not provide secure defenses for other types of users.  
\item {\bf Information Leakage. }The \selpred~algorithm computes on the user's correct password to assign a sequence of predicates to the user. In our mechanisms, $\epsilon$ denotes the level of leakage allowed by the client where smaller values of $\epsilon$ indicate lower levels of leakage. For example, when $\epsilon = 0$, $\selpred$ uses a uniform distribution to assign predicate sequences, but this distribution changes with larger $\epsilon$. 
\end{itemize}
}
\cut{
%\vspace{-0.4cm}
\section{Discussions and Future Work}
%\vspace{-0.4cm}
\label{sec:Discussions and Future Work}
% !TEX root = pwd_system_llncs.tex
\checkpoint
All the \clientcash~ mechanisms examined give reductions in the percentage of passwords breached by an optimal offline adversary. We also make the following recommendations of when a user should use a certain mechanism. 

\begin{itemize}
\item {\bf Two Rounds vs. Three Rounds.} Based on the results in Figure \ref{fig:cmpfig}, we recommend using three rounds of hashing in general. The one possible exception is when the client is using the exponential mechanism with $\epsilon \ge 1.6$. In Figure \ref{subfig:expcmp}, the two round mechanism outperforms the three round mechanism for low values of $\frac{B}{k'|\pwdspace|} \le 0.5$. Considering the trend of three round exponential graphs shown in Figure \ref{subfig:threepred_exp}, this gap between two and three rounds is likely to increase with greater values of $\epsilon$. However, we also note that the three round exponential mechanism achieves greater maximum gains in security and is nonzero for larger values of the adversary budget $B$. We ask that the client considers these when choosing between two and three rounds in this case.   
\end{itemize}

\cut{All the \clientcash~ mechanisms examined give reductions in the percentage of passwords breached by an optimal offline adversary, but given a choice between using two rounds of hashing and three rounds of hashing (using either the exponential or optimal distributions), we recommend using three rounds. Intuitively, one may expect that a larger number of rounds will give greater reductions in the percentage of passwords breached as it implies more possibilities for the adversary to check, but the true nature of the relationship is not clear since (1) $C_{srv}$ is constant, so increasing the number of rounds decreases the value of the key stretching constant $k$, and (2) the adversary knows the distribution of $\hprob{j}$'s and can still optimize an attack based on these values. Nevertheless, using three rounds gives better results both with the optimal distributions (used when the adversary budget can be estimated accurately) and the exponential mechanism (used when the adversary budget has lots of variation or when it cannot be estimated accurately). }

\cut{\anote{insert figures here, not much discussion needed. } In this paper, we present only the results for two and three rounds to show the promise of \clientcash, but the effect of more rounds of hashing should be investigated and may result in greater reductions in passwords breached.}

\subsection{Future Work}

The main goal of this work was to formalize the properties of the \clientcash~mechanism and show that we can obtain significant reductions in the number of passwords breached by an adversary. Naturally, there are several extensions to our work which can increases the reduction in passwords breached even further. 
\begin{itemize}
\item {\bf Number of rounds. }As discussed previously in \textsection \ref{sec:Discussions and Future Work}, we only analyze mechanisms using two and three rounds of hashing. We have shown that using three rounds is always better than using two rounds, and it is possible that this behavior continues by considering more rounds of hashing. 
\item {\bf Predicate structure. }In our mechanisms we set $\ell_i = n$ for all $i \in \{1,\ldots,n\}$, but we can optimize with respect to the $\ell_i$'s as well. More generally, we can use different types of predicates to produce a symmetric $\outputspace$ (see \textsection \ref{sec:sys_properties} and Theorem \ref{thm:symmetric_output}). We note that this change would fundamentally alter our adversary model. 
\item {\bf Non-uniform password selection. }We assume in this work that a user who opts to use \clientcash~is mindful of security and therefore will choose passwords roughly uniformly at random. A typical user does not pick passwords uniformly, which is evident through recent password breaches \anote{cite this}. Thus, we can also design \clientcash~with respect to a particular password distribution over the set of users and treat the formulation in this paper as a special case. 
\end{itemize} }

%\vspace{-0.6cm}
\section{Conclusions and Future Work}
%\vspace{-0.45cm}
\label{sec:Conclusions}
% !TEX root = ASIA_CCS_PWD_SYSTEM_2016.tex
In this work, we have introduced \clientcash~ as a novel client-side key stretching mechanism which gives better security guarantees than traditional key-stretching procedures.We make several contributions. First, we show how to introduce randomness into a client-side key stretching algorithms through the use of  halting predicates which are selected randomly at the time of account creation. Second, we formalize the problem of finding the optimal running time distribution subject to certain cost constraints for the client and certain security constrains on the halting predicates. Finally, we demonstrate that \clientcash~can reduce the adversary's success rate by up to $21\%$ ( in the $n=3$ round case). These results demonstrate the promise of the \clientcash~mechanism. In the future we hope that researchers will extend our analysis of \clientcash~to handle non-uniform password distributions. We also speculate that it may be possible to obtain even larger reductions in the adversary's success rate by using predicate sequences $\outputspace$ with more rounds of hashing $n > 3$. Future work might extend our analysis of \clientcash~ to explore this possibility.

\section*{Acknowledgments}
This work was completed in part while Jeremiah Blocki was visiting the Simons Institute for the Theory of Computing, supported by the Simons Foundation and by the DIMACS/Simons Collaboration in Cryptography through NSF grant $\#$CNS-1523467. Anirudh Sridhar was supported by a grant from the PNC Center for Financial Services Innovation. Any opinions, findings, and conclusions or recommendations expressed in 
this material are those of the author(s) and do not necessarily reflect 
the views of Microsoft, the PNC Center for Financial Services Innovation at 
Carnegie Mellon University or the National Science Foundation.

%\subsection{Two Round Optimization}
%\label{subsec:Two Round Optimization}
%\input{tworound_opt}

%\subsection{Three Round Optimization}
%\label{subsec:Three Round Optimization}
%\input{threeround_opt}

%%%%%%%%%%%%%%%%%%%%%%

{\bibliographystyle{abbrv}
\bibliography{password}}
\fullversion{

\appendix
% !TEX root = ASIA_CCS_PWD_SYSTEM_2016.tex
\cut{\section{Complete List of Notation}
Table \ref{tab:full_notation} summarizes much of the notation used throughout this paper. 
\begin{table}[tbh]
\centering
\begin{tabular}{|c|p{3.5in}|}
\hline
Term & Description \\
\hline
$\hash$ & Cryptographic hash function\\
\hline
$\hash^t$ & Hash function with $t$ iterations\\
\hline
$k$ & The number of hash iterations in each round of \clientcash\\
\hline
$n$ & The maximum number of hashing rounds. $nk$ is the maximum number of hash iterations\\
\hline
$a$ & Account name\\
\hline
$C_{srv}$ & The user's amortized cost for computing the hash of a password. \\
\hline
$C_H$ & The computational cost of one iteration of $\hash$. \\
\hline
$\pwdspace$ & Space of all possible passwords\\
\hline
$\outputspace$ & The set of all valid predicate sequences\\
\hline
$\outputspace_{j,pwd}$ & The set of outcomes with stopping time $j$ given a fixed $pwd \in \pwdspace$.\\
\hline
$O_j$ & The size of $\outputspace_{j,pwd}$ for a symmetric set $\outputspace$. \\
\hline
\cut{$\prob{o}$ & The probability that $o \in \outputspace$ is selected by \selpred for a fixed $pwd$. \\
\hline}
$\prob{j}$ & The probability of each outcome $o \in \outputspace_{j,pwd}$ when $\outputspace$ is symmetric.\\
\hline
$\hprob{j}$ & The probability that a given password $pwd$ has stopping time $j$ when $\outputspace$ is symmetric.\\
\hline
$s_u$ & The random salt for user $u$. \\
\hline
$\stopcond(pwd,o)$ &  The stopping time of $pwd$ given $o \in \outputspace$. The derived hash of the password is $\hash^{\stopcond(pwd,o)\times k}(pwd,s)$. \\
\hline
$s \stackrel{\$}{\gets}\{0,1\}^L$ & The salt $s$ is randomly selected uniformly from $\{0,1\}^L$. \\
\hline
$\pdetB$ & The probability that the adversary breaches a given user account in the traditional password hashing system. \\
\hline
$\padvB$ & The probability that the adversary breaches a given user account with Client CASH. \\
\hline
$\epsilon$ & Differential privacy security constant. \\
\hline
$U$ & Utility functions for the exponential mechanism. \\
\hline

\cut{
$S_m$ & The set of passwords the adversary hashes with $\hash^k$ at least $m$ times. \\
\hline
$T_m$ & The set of passwords with stopping time at most $m$. \\
\hline
$A^c$ & The complement of set $A$. \\
\hline
$\advstrat{i}(S)$ & Given a certain subset $S$ of all possible feasible adversary strategies, this is the collection of strategies in $S$ such that the adversary maximizes the number of passwords hashed $i$ times with $\hash^k$. \\
\hline}
\end{tabular}
\caption{Notation}
\label{tab:full_notation}
\end{table}}

\cut{
\section{Account Creation and Password Derivation}
\subsection{Account Creation} \label{apdx:accountcreation}
The client will run Algorithm \ref{alg:CreateAccount} to setup the account. Algorithm  \ref{alg:CreateAccount} is a randomized algorithm that runs the $\selpred$ protocol to produce $o_u \in \outputspace$, draws a random salt value $s_u$ and stores these values in a table on the client.The client-side function \reproduce (presented as Algorithm \ref{alg:Reproduce}) is then used to derive a password $\hash^{tk}(pwd_u,s_u)$, which is sent to the server. Here, $t = \stopcond(pwd_u,o_u)$ is the stopping time given by the predicates $o_u$ and $s_u$ is the random $L$-bit salt assigned to the user $u$. We emphasize that $o_u$ is selected based on some probability distribution which may depend both on $pwd_u$ and our security parameter $\epsilon$. The account creation protocol is presented as Algorithm 1. We intentionally omit the workings of $\selpred$ for now and treat it as a black box in this setting. In later sections we will show how to construct a randomized algorithm $\selpred$ which minimizes the adversary's success rate subject to certain security and cost constraints which will also be introduced in subsequent sections.

\subsection{Authentication} \label{apdx:authentication}

When the user $u$ attempts to access the account with the password guess $pwd_g \in \pwdspace$, the client first locates the record $(o_u,s_u)$. Then the client-side runs the \reproduce function to derive the password $\hash^{k\cdot\stopcond(pwd_g,o_u)}(pwd_g,s_u)$. The result is sent to the server to be accepted or rejected based on the server's record $(u,H_u)$. Authentication is guaranteed when $pwd_g = pwd_u$ and is very unlikely when $pwd_g \neq pwd_u$ because $\hash$ is collision resistant. The client-side protocol \reproduce is presented as Algorithm \ref{alg:Reproduce}. Unlike Algorithm  \ref{alg:CreateAccount},  Algorithm \ref{alg:Reproduce} is entirely deterministic. We note that the only role of the server is verifying whether the hash $H_g$ generated by the password guess $pwd_g$ matches the value $H_u$ stored in the server. 
}

\section{Missing Proofs}

\begin{remindertheorem}{\ref{thm:symmetric_output}}
Let $n$ be the number of rounds. Suppose that $\outputspace$ is the set of all possible predicate sequences such that for $\{P_1,\ldots,P_n\} \in \outputspace$, $P_m \in \{P_{0,\ell_m},\ldots P_{\ell_m-1,\ell_m}\}$ for some fixed $\ell_m$ for all $m \in \{1,\ldots,n\}$. Then for all $pwd,pwd' \in \pwdspace$, $\outputspace$ is symmetric, i.e. for all $j \in \{1,\ldots,n\}$, $|\outputspace_{j,pwd}| = |\outputspace_{j,pwd'}|$. 
\end{remindertheorem}
\\
\\
\begin{proofof}{Theorem \ref{thm:symmetric_output}}
We proceed by a simple counting argument. Suppose that $j < n$. Then $\outputspace_{j,pwd}$ consists of all outcomes which give a stopping time of $j$, so for every $o = \{P_1,\ldots P_n\} \in \outputspace_{j,pwd}$, we must have $P_i (\hash^{ik}(pwd,s_u)) = 0$ for $i < j$ and $P_j(\hash^{jk}(pwd,s_u)) = 1$. Due to the structure of $\outputspace$, we can directly calculate the value $\left|\outputspace_{j,pwd} \right|$, since for all choices of $pwd$, there will be 1 choice of the $m$th predicate which will give an evaluation of $1$ on the iterated hash, and $\ell_m - 1$ choices that give an evaluation of 0. Thus, for $j < n$, $$
\left|\outputspace_{j,pwd}\right| = (\ell_1-1)\times\ldots (\ell_{j-1}-1)\times 1 \times \ell_{j+1} \times \ldots \times \ell_{n-1}
$$
and when $j = n$, since we have a sequence of $n-1$ predicates, 
$$
\outputspace_{n,pwd} = (\ell_1-1)\times \ldots \times (\ell_{n-1}-1)
$$
Since these calculations are independent of the password selected we can set $\outputspace_j \doteq (\ell_1-1)\times\ldots (\ell_{j-1}-1)\times 1 \times \ell_{j+1} \times \ldots \ell_{n-1}$ for $j < n$ and $\outputspace_n = (\ell_1-1)\times \ldots (\ell_{n-1}-1)$. Now $
\left|\outputspace_{j,pwd}\right| = \outputspace_j$ for any password $pwd \in \pwdspace$ and any $j \in \{1,\ldots,n\}$. Thus, $\outputspace$ is indeed symmetric. 
\end{proofof}

Theorem \ref{thm:diff_privacy}, stated informally in Table \ref{tab:Constraints}, shows how we can simplify our security constraints using the symmetric set $\outputspace$ from Theorem \ref{thm:symmetric_output}. The key observation is that Theorem \ref{thm:diff_privacy} gives us one constraint for each $i,j \in \{1,\ldots,n\}$ instead of multiple constraints for each password pair of passwords $pwd\ne pwd'$ that the user might select. While the space of passwords may be very large, $n$, the number of rounds of hashing, will typically be quite small (e.g., in this paper $n \in \{2,3\}$). 

\newcommand{\thmdiffprivacy}{
Suppose that $\outputspace$ is constructed as specified in Theorem \ref{thm:symmetric_output}, equation \ref{eq:symmetry} holds and $\forall i,j \in \{1,\ldots,n\}, \frac{\prob{i}}{\prob{j}} \le e^\epsilon$. Then $\forall pwd,pwd' \in \pwdspace, \forall o \in \outputspace, \frac{ \Pr[\selpred(pwd) = o]}{\Pr[\selpred(pwd') = o]} \le e^\epsilon$. 
}

\begin{theorem}
\label{thm:diff_privacy}
\thmdiffprivacy
\end{theorem}

\begin{proofof}{Theorem \ref{thm:diff_privacy}}
We begin by noting that for all passwords $pwd \in \pwdspace$ and for all outcomes $o \in \outputspace$ we can find some $m \in \{1,\ldots,n\}$ such that  \[ \Pr[\selpred(pwd) = o] = \prob{m} \ , \] because $\{\outputspace_{j,pwd}\}$ partitions $\outputspace$. Let $pwd,pwd' \in \pwdspace$ be arbitrarily selected. Then for some $i,j \in \{1,\ldots,n\}$ we have $$
\frac{\Pr[\selpred(pwd) = o]}{\Pr[\selpred(pwd') = o]} = \frac{\prob{i}}{\prob{j}}
$$
which is in turn less than or equal to $e^\epsilon$ by our second assumption. Thus, $$
\frac{\Pr[\selpred(pwd) = o]}{\Pr[\selpred(pwd') = o]} \le e^\epsilon
$$
for all $pwd,pwd' \in \pwdspace$ and for all $o \in \outputspace$. Thus $\epsilon$-differential privacy is satisfied by these conditions. 
\end{proofof}

Similarly, Theorem \ref{thm:cost_constraints},stated informally in Table \ref{tab:Constraints}, says that we exploit symmetry to simplify our cost constraint. Before we had to satisfy separate cost constraints for all passwords $pwd \in \pwdspace$. With symmetry we only have to satisfy one cost constraint.

\newcommand{\thmcostconstraints}{
Suppose that $\outputspace$ and the $\prob{i}$'s are constructed as specified in Theorems \ref{thm:symmetric_output} and \ref{thm:diff_privacy}, and that $\sum\limits_{i=1}^n i \cdot O_i \prob{i} \le \alphak$. Then for all $pwd \in \pwdspace$ we have $\sum\limits_{i=1}^n i \cdot \hprob{i,pwd}  \le \alphak$. 
}

\begin{theorem}
\label{thm:cost_constraints}
\thmcostconstraints
\end{theorem}

\begin{proofof}{Theorem \ref{thm:cost_constraints}}
Recall that $E_j$ is the event that $\selpred(pwd) \in \outputspace_{j,pwd}$ for a fixed $pwd \in \pwdspace$. With the symmetric construction of $\outputspace$ and the design of the probability distribution, we have that $\hprob{j} = $
\begin{equation*}
\begin{split}
 \sum\limits_{o \in \outputspace_{j,pwd}} \Pr[\selpred(pwd) = o]
& = \sum\limits_{o \in \outputspace_{j,pwd}} \prob{j} \le O_j \prob{j} \ .
\end{split}
\end{equation*}
Thus, $$
\sum\limits_{j=1}^n j\times O_j\prob{j} = \sum\limits_{j=1}^n j \times \hprob{j} \le \alphak
$$

\end{proofof}

\cut{\begin{remindertheorem}{\ref{thm:diff_privacy}}
\thmdiffprivacy
\end{remindertheorem}\\}

We defer the proof of Theorem \ref{thm:fullpadv} until appendix \textsection \ref{apdx:Adversary}  as it is more involved.

% Moved this proof back to main body.
\cut{
\begin{remindertheorem}{\ref{thm:diffPrivacy2}}
\thmDiffPrivacyTwo
\end{remindertheorem}
\begin{proofof}{Theorem \ref{thm:diffPrivacy2}}
Let $w_{o,pwd} \doteq e^{\epsilon U(pwd,o)}$ denote the weight of outcome $o \in \outputspace$ given password $pwd$ and let $W_{pwd} \doteq \sum\limits_{o \in \outputspace} w_{o,pwd}$ denote the cumulative weight of all $o \in \outputspace$. Because $\outputspace$ is symmetric we have $W_{pwd} = W_{pwd'}$ for any pair of passwords $pwd,pwd' \in \pwdspace$. Thus, 
\begin{eqnarray*}
 \frac{\Pr\left[\mathbf{SelectPredicate}\left(pwd\right)=o\right]}{\Pr\left[\mathbf{SelectPredicate}\left(pwd'\right)=o\right]} 
&=& \frac{w_{o,pwd}/W_{pwd}}{w_{o,pwd'}/W_{pwd'}} \\  =  \frac{w_{o,pwd}}{w_{o,pwd'}} 
\leq e^{\epsilon\left|U(pwd',0)-U(pwd,o) \right|} &\leq& e^\epsilon \ .
\end{eqnarray*}
\end{proofof}}

\cut{
\begin{remindertheorem}{\ref{thm:cost_constraints}}
\thmcostconstraints
\end{remindertheorem}
}

\subsection{Future Work}

The main goal of this work was to formalize the properties of the \clientcash~mechanism and show that we can obtain significant reductions in the number of passwords breached by an adversary. Naturally, there are several extensions to our work which can increases the reduction in passwords breached even further. 
\begin{itemize}
\item {\bf Number of rounds. } We only analyze mechanisms using two and three rounds of hashing. We have shown that using three rounds is always better than using two rounds, and it is possible that this behavior continues by considering more rounds of hashing. 
\item {\bf Predicate structure. }In our mechanisms we set $\ell_i = n$ for all $i \in \{1,\ldots,n\}$, but we can optimize with respect to the $\ell_i$'s as well. More generally, we can use different types of predicates to produce a symmetric $\outputspace$ (see \textsection \ref{sec:sys_properties} and Theorem \ref{thm:symmetric_output}). We note that this change would fundamentally alter our adversary model. 
\item {\bf Non-uniform password selection. }We assume in this work that a user who opts to use \clientcash~is mindful of security and therefore will choose passwords roughly uniformly at random. A typical user does not pick passwords uniformly, which is evident through recent password breaches. Thus, we can also design \clientcash~with respect to a particular password distribution over the set of users and treat the formulation in this paper as a special case. 
\end{itemize}

% !TEX root = ASIA_CCS_PWD_SYSTEM_2016.tex
% 
% 
\cut{\section{Extra Plots: $n=2$ vs $n=3$ rounds}
Figures \ref{subfig:expcmp} and \ref{subfig:optcmp} compare the two-round \clientcash~mechanism mechanism to the three-round \clientcash~mechanism. In general, the three-round mechanism outperforms the two-round mechanism. This suggests that we may be able to obtain even greater reductions $G_B$ in the adversary's success rate by moving to a four round mechanism $n=4$. However, we note that larger values of $n$ will not necessarily be superior due to our differential privacy constraint. For example, the differential privacy constraint ensures outcomes with stopping time $\stopcond(pwd,o)=n$ will be output with comparable probability to outcomes $o'$ with short stopping times.

\begin{figure*}[t!]
    \centering
    \begin{subfigure}[t]{0.5\textwidth}
        \centering
        \includegraphics[width=1.1\textwidth]{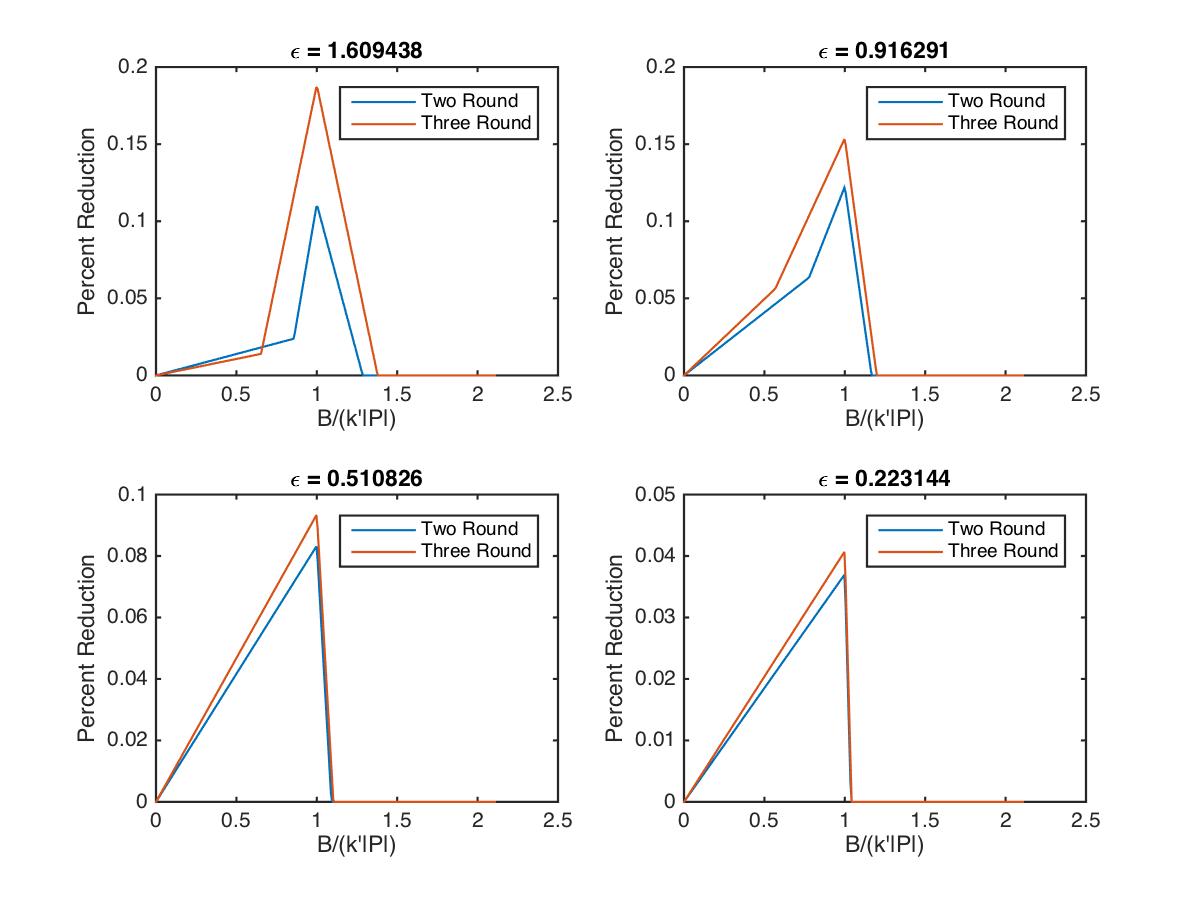}
        \caption{Comparison of exponential mechanisms. }
        \label{subfig:expcmp}
    \end{subfigure}%
    ~ 
    \begin{subfigure}[t]{0.5\textwidth}
        \centering
        \includegraphics[width=1.1\textwidth]{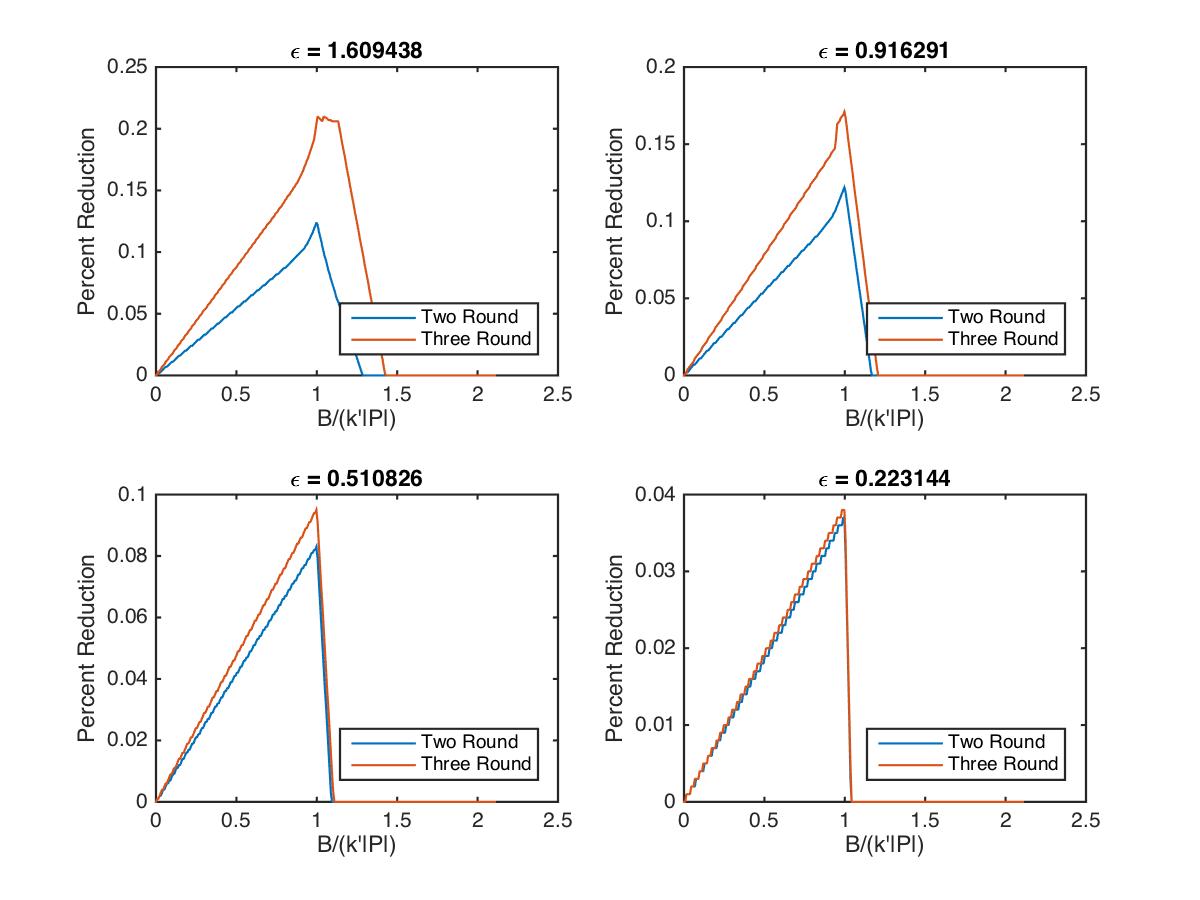}
        \caption{Comparison of optimal mechanisms. }
        \label{subfig:optcmp}
    \end{subfigure}
    \caption{
    Comparison of the percent reduction in passwords between two-round mechanisms (blue) and three-round mechanisms (orange) as a function of $B$ with a normalizing factor of $\frac{1}{k'|\pwdspace|} = \frac{C_H}{C_{srv}|\pwdspace|}$ for various values of $\epsilon$. The results for the exponential mechanism are shown in (a), while the optimal mechanism is shown in (b).  }
\label{fig:cmpfig}
\end{figure*}}

\section{Adversary Analysis} \label{apdx:Adversary}
% !TEX root = ASIA_CCS_PWD_SYSTEM_2016.tex
Recall that $S_m$ denotes the set of passwords hashed at least $m$ times by the adversary, and $T_m$ denotes the set of passwords in $\pwdspace$ with stopping time at most $m$. Formally, 

\begin{equation*}
\begin{split}
S_m & = \{pwd \in \pwdspace: \hash^{m\times k}(pwd) \text{ is computed by the adversary}\}\\
T_m & = \{pwd \in \pwdspace: \upred{m}(\hash^{m\times k}(pwd)) = 1\} \text{ for } m < n\\
\end{split}
\end{equation*}
Recall that the \textbf{strategy} of the adversary is a n-tuple of sets $(S_1,S_2,...S_n)$ such that 
\begin{itemize}
\item $S_m$ is the set of passwords hashed at least $m$ times
\item $S_1 \subseteq \pwdspace$
\item $S_p \subseteq S_{p-1}\cap T_{p-1}^c$ for $p > 1$
\item $\sum\limits_{i=1}^n |S_i| = \frac{B}{k}$
\end{itemize}
We also assume that for $m \le n$, $|S_m|$ is either 0 or of a non-negligible quantity compared to the number of iterated hashes the adversary can compute, $\frac{B}{k}$. Since users pick passwords uniformly at random, if the size of $S_m$ is negligible compared to $\frac{B}{k}$, then the increase in the adversary's probability of success will also be negligible.\\
\\
Let $X \subseteq \pwdspace$ be of non-negligible size and suppose that $\upred{i} = P_{t,\ell_i}$ for some $t < \ell_i$. Treating our cryptographic hash like a random oracle and assuming that $|\pwdspace| \gg n$, we have with high probability that $$
\left| \Pr[pwd \in T_i\mid pwd \in X] - \frac{1}{\ell_i} \right|  = \left| \frac{|T_i \cap X|}{|X|} - \frac{1}{\ell_i}  \right|< \epsilon_i
$$
for small $\epsilon_i > 0$. We thus assume for simplicity that $\ell_i \times |T_i \cap X| = |X|$ for all $i \in \{1,\ldots,n-1\}$ as long as $X$ is of non-negligible size. Some cases of interest are when $X = \pwdspace$ to obtain $\ell_i \times |T_i| = |\pwdspace|$ and $X = S_i$ to obtain $\ell_i \times |S_i \cap T_i| = |S_i|$.  \\
\\
For $m < n$ define $$
B_m := |\{ pwd \in S_m : \stopcond(pwd,o_u) = m \}|
$$
where $o_u$ is the user's output from the \selpred function. We also define $B_n := |S_n|$. Since we assume we face an optimal adversary, each $pwd \in S_m$ such that $\stopcond(pwd,o_u) = m$ will also satisfy $pwd \in T_m$ for $m < n$. Thus, $B_m = |S_m \cap T_m|$ for $m < n$. \\
\\
From this formulation, we obtain the following constraints
\begin{itemize}
\item $S_1 \subseteq \pwdspace \Rightarrow \ell_1 B_1 \le |\pwdspace|$
\item $|S_m| \le |S_{m-1} \cap T_{m-1}^c| \Rightarrow \ell_m B_m \le (\ell_{m-1}-1)B_{m-1}$ \\for $1 < m \le n$
\item $|S_n| \le |S_{n-1} \cap T_{n-1}^c| \Rightarrow B_n \le (\ell_{n-1}-1)B_{n-1}$
\item $\sum\limits_{i=1}^n |S_i| = \frac{B}{k} \Rightarrow \sum\limits_{i=1}^{n-1}\ell_i B_i + B_n = \frac{B}{k}$
\end{itemize}
We can rephrase these constraints by normalizing the values of the $B_i$'s relative to $\frac{B}{k}$. Define$$
b_m = \frac{|S_m|}{B/k} = 
\begin{cases}
\frac{k\ell_mB_m}{B} &\text{if } m < n\\
\frac{kB_m}{B} &\text{if } m = n
\end{cases}
$$
Intuitively, $b_m$ represents the fraction of the budget spent on hashing passwords at least $m$ times. We additionally note that since each of the $B_i$'s must be less than $\frac{B}{k}$, each of the $b_i$'s must be less than or equal to 1. Then we have
\begin{itemize}
\item $\mathbf{b} = (b_1,\ldots,b_n) \in [0,1]^n$
\item $b_1 \le \invpwdratio \Rightarrow b_1 \le \min\left\{1,\invpwdratio\right\}$
\item $b_m \le \frac{\ell_{m-1}-1}{\ell_{m-1}}b_{m-1} \text{ for } 1 < m \le n$
\item $\sum_{i=1}^n b_i = 1$
\end{itemize}
These are the necessary conditions for a feasible (optimal) adversary strategy. We can also show that is it sufficient as well, that every $\mathbf{b}$ that satisfies these properties corresponds to an adversary strategy since the conversion process from the adversary strategy tuple of sets $(S_1,\ldots,S_n)$ to $\mathbf{b}$ is a reversible process (the usage of $\mathbf{b}$ gives a set size, and $(S_1,\ldots,S_n)$ gives the sets themselves, but the contents do not matter much since passwords are selected uniformly at random). We can therefore equivalently define the feasible strategy region $F \subseteq \left[0,\min\left\{1,\invpwdratio\right\}\right]^n$ such that all $\mathbf{b} \in F$ satisfy the above properties. 

\subsection{Maximum Adversary Budget}

If the adversary's budget is sufficiently large, then the adversary will almost certainly be able to retrieve the password and number of hashes necessary to breach an account. For example, if $\frac{B}{k} = n|\pwdspace|$, the adversary would be able to hash every password $n$ times and therefore gain access to the account. We wish to find the minimum budget size such that the adversary succeeds with probability 1. 

\textbf{Step 1: } The adversary determines and checks the set of passwords such that $\upred{1}(\hash_k(pwd)) = 1$. The adversary must hash every password in the space, so there are $|\pwdspace| \times k$ computations. 

\textbf{Step m $(m < n)$: } The adversary determines and checks the set of remaining passwords such that $\upred{m}(\hash_k^{m}(pwd)) = 1$. There are roughly $\prod\limits_{i=1}^{m-1} \left( \frac{\ell_i-1}{\ell_i} \right)|\pwdspace|$ candidate passwords, so there are $\prod\limits_{i=1}^{m-1} \left( \frac{\ell_i-1}{\ell_i} \right)|\pwdspace| \times k$ computations.

\textbf{Step n: } The adversary determines and checks the set of passwords such that $\stopcond(pwd,o_u) = n$. These are the leftover passwords, so the adversary must hash all of these to breach the account. Thus there are $\prod\limits_{i=1}^{n-1} \left( \frac{\ell_i-1}{\ell_i} \right)|\pwdspace| \times k$ computations.

In total, for the adversary to succeed with probability 1, 
\begin{equation*}
\begin{split}
B  &\ge |\pwdspace|\times k + \sum\limits_{m=1}^{n-1} \prod\limits_{i=1}^{m} \left( \frac{\ell_i-1}{\ell_i} \right)|\pwdspace| \times k\\
\end{split}
\end{equation*}
We thus assume in the rest of our analysis that $$
\frac{B}{k} < \left( 1 + \sum\limits_{m=1}^{n-1} \prod\limits_{i=1}^{m} \left( \frac{\ell_i-1}{\ell_i} \right) \right) |\pwdspace|
$$
It is of interest to analyze the case where $\ell_i = n$ for all $i$ when we have $n$ rounds. Then $\frac{B}{k} < $
\begin{equation*}
\begin{split}
\left( 1 + \sum\limits_{m=1}^{n-1} \left( \frac{n-1}{n} \right)^m  \right)|\pwdspace| & = \left( \sum\limits_{m=0}^{n-1} \left( \frac{n-1}{n} \right)^m \right) |\pwdspace|\\
& = \left( \frac{n^n - (n-1)^n}{n^{n-1}}  \right) |\pwdspace|
\end{split}
\end{equation*}
This is clearly an increasing function in $n$, so as long as we pick sufficient $n \ll |\pwdspace|$ we can make sure the adversary's probability of success is less than 1. 

\subsection{The Adversary's Probability of Success}

Suppose that the adversary uses strategy $\mathbf{b} \in F$. We define $P_b$ to be the adversary's probability of success with this strategy. Then $P_b$ is simply the probability of selecting the right password and hashing it the correct number of times. Suppose that the user's correct password is $pwd_u = pwd^*$ for user $u$, and that $\stopcond(pwd^*,o_u) = i$. Then the adversary's probability of success is the same as the probability that $pwd^* \in S_i$, given that it has not reached its stopping time yet. Recall that 
$$
T_m = \{pwd \in \pwdspace: \upred{m}(\hash^{m\times k}(pwd)) = 1\} \text{ for } m < n
$$
Then, for $i > 1$, $\Pr[\text{Adversary succeeds} \cap \stopcond(pwd^*,o_u) = i] = $
\begin{equation*}
\begin{split}
& \Pr[pwd_u = pwd^*] \times \hprob{i}\times\Pr\left[pwd^* \in S_i \mid pwd^* \in  \bigcap\limits_{j=1}^{i-1}T_j^c\right]\\
 =& \frac{1}{|\pwdspace|} \hprob{i} \frac{|S_i|/|\pwdspace|}{|\bigcap\limits_{j=1}^{i-1}T_j^c|/|\pwdspace|} = \frac{1}{|\pwdspace|}\hprob{i}\left(\frac{\frac{b_i\cdot B}{k}}{\prod\limits_{j=1}^{i-1} \left(\frac{\ell_j-1}{\ell_j}\right)}\right) \\
 =& \pwdratio \hprob{i} b_i \prod\limits_{j=1}^{i-1} \frac{\ell_j}{\ell_j-1}\\
\end{split}
\end{equation*}
In the case where $i=1$, the probability is simply $\pwdratio \hprob{1}b_1$ since the adversary has no information about stopping times before hashing passwords for the first time. We can now calculate $P_b$ to be
\begin{equation*}
\begin{split}
P_b & = \sum\limits_{i=1}^n \Pr[\text{Adversary succeeds} \cap \stopcond(pwd^*,o_u) = i]\\
& = \pwdratio \left( \hprob{1}b_1 + \sum\limits_{i=2}^n \hprob{i}b_i \prod\limits_{j=1}^{i-1}\frac{\ell_j}{\ell_j - 1}   \right)
\end{split}
\end{equation*}
However, the adversary has control over the particular strategy to execute. Let $\padv$ denote the probability of success for an optimal adversary. Then $\padvB = \max\limits_{\mathbf{b} \in F} P_b =$ $$
   \pwdratio \max\limits_{\mathbf{b} \in F} \left\{\hprob{1}b_1 + \sum\limits_{i=2}^n \hprob{i}b_i \prod\limits_{j=1}^{i-1}\left(\frac{\ell_j}{\ell_j - 1}  \right) \right\} \ .
$$
As an interesting special case of importance to us, let $\ell_i = n$ for all $i \le n$. Then 
\begin{equation*}
\begin{split}
\padvB & = \pwdratio \max\limits_{\mathbf{b} \in F} \left\{  \hprob{1}b_1 + \sum\limits_{i=2}^n \hprob{i}b_i \left(  \frac{n}{n-1}  \right)^{i-1}   \right\}\\
& = \pwdratio \max\limits_{\mathbf{b} \in F} \left\{  \sum\limits_{i=1}^n \hprob{i}b_i \left( \frac{n}{n-1}  \right)^{i-1}   \right\}
\end{split}
\end{equation*}

%\subsection{Optimal Adversary Strategies}
%\input{adv_opt}

\subsection{Dominant Adversary Strategies}
% !TEX root = ASIA_CCS_PWD_SYSTEM_2016.tex
Calculating the optimal $\padv$ for fixed $\hprob{j}$'s may be made efficient if we consider \textit{dominant strategies} of the adversary; i.e. $\mathbf{b^*} \in F$ which yield greater values of $\padv$ than other $\mathbf{b} \in F$ independent of the values of $\hprob{j}$'s. Let $\advstrat{i} \subseteq F$ denote the collection of members of $F$ such that $b_i$ is maximized. Formally, we define $
\advstrat{i} := \{\mathbf{b} \in F: \forall \mathbf{b'} \in F, b_i \ge b'_i\}
$.
Then we claim that the collection $F^* := \bigcup\limits_{i=1}^n \advstrat{i}$ is the collection of dominant strategies of the adversary. 

\begin{remindertheorem}{\ref{thm:fullpadv}}
\thmfullpadv
\end{remindertheorem}

\begin{proofof}{Theorem \ref{thm:fullpadv}}
Recall that the feasible strategy region $F \subseteq [0,1]^n$ for a system with up to $n$ rounds of hashing is defined as follows. For all $\mathbf{b} \in F$, 
\begin{eqnarray*}
&\bullet ~ b_1 \le \min\left\{ 1,\invpwdratio   \right\} ~~~~~~~~~~~~~~~~~~& \bullet~\sum\limits_{i=1}^n b_i = 1\\
&\bullet ~b_m \le \frac{\ell_{m-1}-1}{\ell_{m-1}}b_{m-1}~~~~~\mbox{for $m > 1$. } &
\end{eqnarray*}

and that $$
\padvB = \padvformula .
$$
Noting that $f(\mathbf{b}) = \padvfunction$ is linear in $\mathbf{b}$, we can rewrite the formula for $\padvB$ as a linear optimization problem over the feasible strategy region $F$. This is shown in Optimization Goal \ref{goal:linprog}. We note that the maximum does indeed exist since it can be shown that the feasible region $F$ is closed and bounded. 

\begin{goal}
\centering
\begin{algorithmic}
\State {\bf Input Parameters: } $B,k,n,\pwdspace,\hprob{j},\ell_j$
\State{\bf Variables: } $\mathbf{b} = (b_1,\ldots,b_n)$
\State {\bf maximize } $\padvB = \pwdratio f(\mathbf{b})$ subject to
\State $(\text{Constraint } 1)~ b_1 \le \min\left\{ 1,\invpwdratio   \right\}$ 
\State $(\text{Constraint }m) ~b_m \le \frac{\ell_{m-1}-1}{\ell_m}b_{m-1}$ for $1 < m < n$
\State $(\text{Constraint }n) ~0 \le b_n \le \frac{\ell_{n-1}-1}{\ell_n}b_{n-1}$
\State $(\text{Constraint }n+1)~ b_1 + \ldots + b_n = 1$
\end{algorithmic}
\vspace{-0.2cm}
\caption{The adversary's probability of success as the solution to a linear program. }
\label{goal:linprog}
\end{goal}
It is a well known fact that for linear optimization, maxima are attained at the vertices of the feasible region; that is, for a $m$-variable linear optimization, the maxima occur at the intersection of $m$ constraints (provided we have at least this many constraints). We show that all such maxima of Optimization Goal \ref{goal:linprog} are also elements of $F_B^* = \bigcup\limits_{i=1}^n \advstrat{i}$. 

We first observe that Optimization Goal \ref{goal:linprog} is effectively an optimization in $n-1$ variables; once the values of $n-1$ variables are determined, the last one can be calculated using Constraint $n+1$. In particular, we can write $b_n = 1 - \big(b_1+\ldots+b_{n-1} \big)$ and Constraint $n+1$ would be rewritten as $
\big(b_1+\ldots+b_{n-1} \big) \le 1$ to reflect this substitution. 

We can find the set of possible maxima by considering all possible intersections of $n-1$ constraints. Since there are $n+1$ constraints in total, this amounts to choosing $2$ constraints to exclude in order to determine a possible maxima.

{\bf Case 1: Constraint 1 is selected. } If Constraint 1 is selected, the adversary sets $b_1$ to its maximum possible value. In this case, $\vec{b} \in \advstrat{1} \subseteq F_B^*$. 

{\bf Case 2: Constraint 1 is excluded. } Constraint 1 is one of two constraints that are excluded, so suppose we also exclude Constraint $j$. Suppose $j = n+1$. Then Constraints 2-$n$ are all set to equality, and are thus directly proportional to $b_1$. In particular, $b_n$ is proportional to $b_1$. Thus, if the adversary maximizes $b_1$, $b_n$ is maximized as well. Thus, $\vec{b} \in \advstrat{n} \subseteq F_B^*$. 

 Otherwise, suppose $j \le n$. Then the values of all variables are dependent on $b_1$ and $b_j$. In particular, since all other constraints are set to equality, we have that 
\begin{equation*}
\begin{split}
b_2 & = \frac{\ell_1 - 1}{\ell_2}b_1 ~~~~~
b_3  = \frac{\ell_2-1}{\ell_3}b_2 = \frac{\ell_2-1}{\ell_3}\frac{\ell_1-1}{\ell_2}b_1\\
b_m & = \frac{\ell_{m-1}-1}{\ell_m}b_{m-1} = \prod\limits_{i=2}^{m} \left( \frac{\ell_{i-1} - 1}{\ell_{i}} \right)b_1 \text{ for } m < j
\end{split}
\end{equation*}
and similarily, 
\begin{equation*}
\begin{split}
b_m = \prod\limits_{i=j+1}^m \left(  \frac{\ell_{i-1}-1}{\ell_i}  \right)b_j \text{ for } m > j
\end{split}
\end{equation*}
We can now reduce our linear program to a much smaller set of constraints: 
\begin{equation*}
\begin{split}
 \bullet ~b_1 &\le \min\left\{1,\invpwdratio \right\}~~~\bullet~\sum\limits_{i=1}^n b_i = 1 \\
\bullet~b_j &\le \frac{\ell_{j-1}-1}{\ell_j}b_{j-1} = \prod\limits_{i=2}^{j}\left( \frac{\ell_{i-1} - 1}{\ell_{i}} \right)b_1
\end{split}
\end{equation*}

However, this new feasible system (which has the same objective function as the original linear program) has only 2 variables; each $b_i$ is directly proportional to either $b_1$ or $b_j$, so for some $\alpha,\beta > 0$, $\sum\limits_{i=1}^n b_i = \alpha b_1 + \beta b_j = 1 \Rightarrow b_1 = \frac{1-\beta b_j}{\alpha}$. Thus, we can reduce the system to just one variable. To calculate the maximum of this new system, we need only pick one constraint to set to equality. If we pick the first to set to equality, then $b_1$ is maximized, so by Case 1, $\vec{b} \in \advstrat{1} \subseteq F_B^*$. Otherwise, if we set the second constraint to equality, $b_j$ is equal to its upper bound, which is proportional to $b_1$. By maximizing $b_1$, we thus maximize $b_j$ so $\vec{b} \in \advstrat{j} \subseteq F_B^*$. 

Thus since every potential maximum over the feasible region $F_B$ is in $F_B^* \subseteq F$, 
\begin{equation*}
\begin{split}
\padv & = \padvformula\\
& = \pwdratio \max\limits_{\mathbf{b} \in F_B^*} \left\{   \padvfunction \right\}
\end{split}
\end{equation*}
\end{proofof}
}

\end{document}